\documentclass[aps,prl,twocolumn,showpacs,superscriptaddress,citeauthorscript]{revtex4-2} 

\usepackage{graphicx}
\usepackage{amssymb}
\usepackage{amsmath}
\usepackage{gensymb}
\usepackage{xcolor}
\usepackage{hyperref}
\usepackage[T1]{fontenc}
\usepackage{comment}
\usepackage{lipsum}
\usepackage[normalem]{ulem}

\begin{document}

%--------------------------------------------------------------
\title{Kondo cloud conductance in cavity-coupled quantum dots with asymmetric barriers}
%--------------------------------------------------------------

\author{D. Fossion}
\email{E-mail: diego.fossion@uclouvain.be}
\affiliation{Univ. Grenoble Alpes, CNRS, Grenoble INP, Institut Néel, Grenoble, France}
\affiliation{Univ. Grenoble Alpes, CEA, Grenoble INP, IRIG-Pheliqs, Grenoble, France}
\author{V. Champain}
\affiliation{Univ. Grenoble Alpes, CEA, Grenoble INP, IRIG-Pheliqs, Grenoble, France}
\author{S. Mohapatra}
\author{A. Cavanna}
\author{U. Gennser}
\author{D. Mailly}
\affiliation{Univ. Paris-Saclay, CNRS, Centre de Nanosciences et de Nanotechnologies, Palaiseau, France}
\author{B. Hackens}
\affiliation{Institute of Condensed Matter and Nanoscience, UCLouvain, Louvain-la-Neuve, Belgium }
\author{L. Jansen}
\author{X. Jehl}
\author{S. De Franceschi}
\author{B. Brun-Barrière}
\affiliation{Univ. Grenoble Alpes, CEA, Grenoble INP, IRIG-Pheliqs, Grenoble, France}
\author{H. Sellier}
\email{E-mail: hermann.sellier@neel.cnrs.fr}
\affiliation{Univ. Grenoble Alpes, CNRS, Grenoble INP, Institut Néel, Grenoble, France}

\date{\today}
\begin{abstract}
The Kondo effect emerges when a localized spin is screened by conduction electrons, giving rise to a strongly-correlated many-body ground state. In this work, we investigate this phenomenon in a GaAs/AlGaAs quantum dot, focusing on the spatial extension of the Kondo screening cloud in the electron reservoirs. To probe its properties, the dot is coupled to an electronic Fabry-Pérot interferometer, enabling controlled modulation of the density of states at the Fermi level. The observation of Kondo temperature oscillations indicates a Kondo screening length comparable to the cavity size. Furthermore, we explore how the coupling asymmetry with the two reservoirs affects both the amplitude and the phase of the conductance oscillations, revealing a subtle interplay between coherent transport and Kondo effect. 
\end{abstract}
\maketitle

%--------------------------------------------------------------
%\section{Introduction}
%--------------------------------------------------------------

The Kondo effect is a paradigmatic many-body phenomenon in condensed matter physics \cite{kondo_resistance_1964}. 
It has served as test bench for many-body theories and has played a key role in the understanding of quantum materials such as heavy-fermion compounds \cite{coleman_heavy_2015}.
The Kondo effect exhibits universal scaling laws involving a single characteristic energy scale called the Kondo temperature $T_{\rm K}$ \cite{nozieres_kondo_1974}.
It was originally discovered in metals containing magnetic impurities, where the Kondo effect leads to the screening of the localized impurity spins by the conduction electrons at temperatures well below $T_{\rm K}$ \cite{wilson_renormalization_1975}.
Each impurity forms a many-body singlet state with a surrounding cloud of strongly-correlated electrons, known as the Kondo cloud.
This cloud extends on a characteristic length scale $\xi_{\rm K} = \hbar v_{\rm F} / k_{\rm B} T_{\rm K}$ where $v_{\rm F}$ is the Fermi velocity.

The Kondo effect was also found to occur in semiconductor quantum dots (QD) hosting an odd number of electrons, which can be seen as artificial magnetic impurities with tunable electronic states \cite{goldhaber_gordon_kondo_1998,cronenwett_tunable_1998}.
In this case, the leads connected to the QD act as tunnel-coupled electron reservoirs, providing the required spin screening through spin-flip cotunneling processes.
Similarly to the case of bulk metals, a Kondo cloud is expected to form around the QD, extending into its leads on the length $\xi_{\rm K}$.
With a typical Fermi velocity around 10$^5$~m/s and a Kondo temperature of about 1~K, the Kondo cloud should extend far away from the QD, on a characteristic distance of about 1~micrometer.

The spatial extension of the Kondo cloud has been extensively studied theoretically \cite{sorensen_scaling_1996, barzykin_kondo_1996, borda_kondo_2007, affleck_friedel_2008, holzner_kondo_2009, busser_numerical_2010, lee_macroscopic_2015, snyman_robust_2015}, but its experimental observation remains a major challenge. 
In the case of magnetic impurities on conducting surfaces, probing the local density of states by scanning tunneling microscopy has provided direct evidence of the Kondo effect \cite{manoharan_quantum_2000, pruser_long-range_2011}, but the observed features were limited to distances much shorter than the expected Kondo length.

For Kondo clouds around semiconductor QDs, several approaches have been proposed.
In particular, finite-size reservoirs with dimensions comparable to the Kondo length should exhibit Kondo temperature oscillations, since resonances in the density of states enhance the Kondo screening process at the Fermi level \cite{affleck_detecting_2001, simon_transport_2006, simon_kondo_2003, pereira_kondo_2008, patton_probing_2009, park_how_2013}. 
Experimental evidence of this prediction has been reported in the groundbreaking work of Ref.~\cite{borzenets_observation_2020}, indicating a Kondo length of micrometer size consistent with the expectation. 
Intriguingly, the Kondo conductance was found to oscillate out of phase with $T_{\rm K}$. 
This property, which was not investigated further in that work, is the main focus of the present letter.

Here we evidence the critical role of barrier asymmetry in the conductance oscillations of a Kondo QD coupled to a finite-length reservoir exhibiting Fabry-Pérot (FP) resonances. 
While the oscillations of $T_{\rm K}$ are consistently in phase with the cavity modes, the Kondo conductance exhibits either in-phase or out-of-phase oscillations, depending on the relative coupling of the QD to the cavity and to the opposite reservoir.
We attribute this surprising result to the crucial role played by the barrier asymmetry in achieving perfect transmission through the Kondo peak at the Fermi level. 
This symmetry condition is a general property of resonant tunneling through a quantum state, but this state is here of many-body origin, as it is a Kondo resonance.

We also pinpoint the issue with measuring the Kondo temperature of an hybrid dot-cavity system through the usual temperature dependence of the conductance, since this dependence is strongly affected by the structured density of states of the cavity.
Our results rather rely on the bias dependence of the conductance as a more accurate probe of the Kondo energy scale $T_{\rm K}$.

\begin{figure*}
    \centering    
    \includegraphics[width=\textwidth]{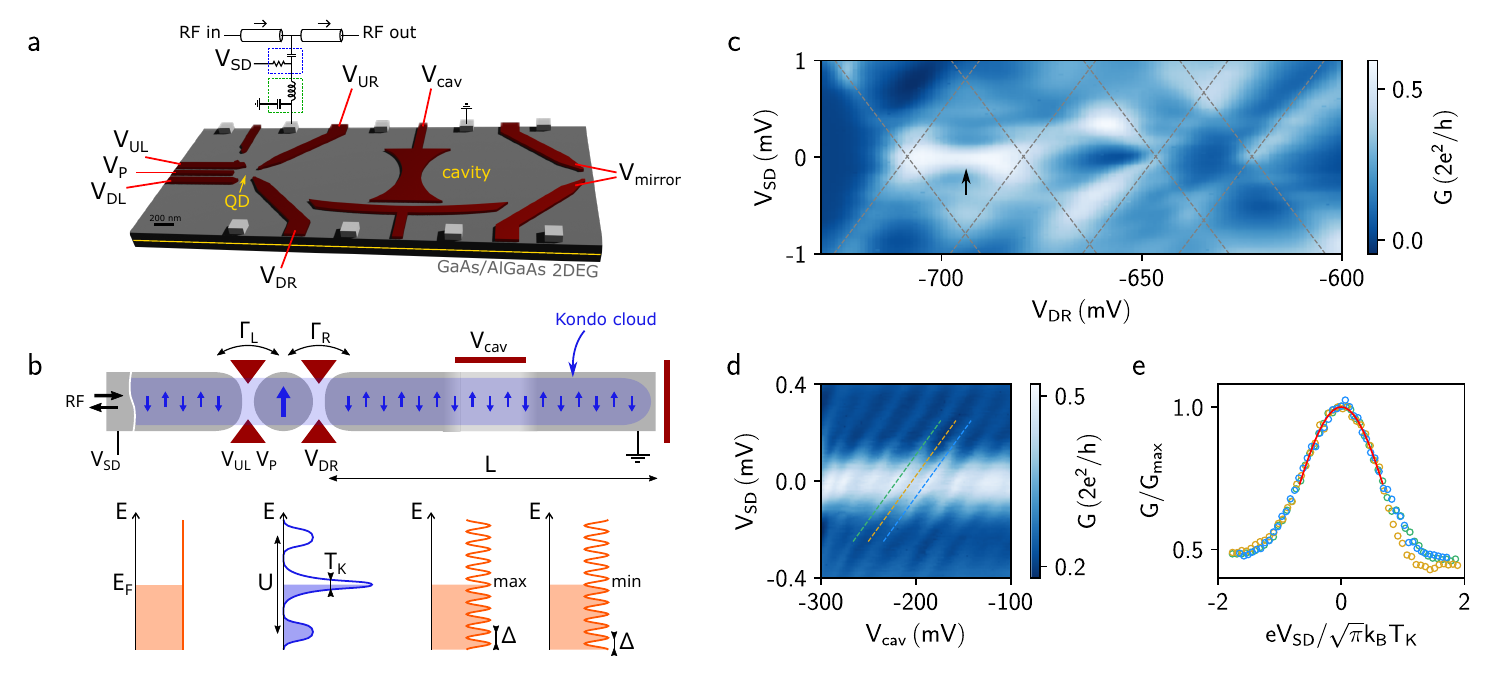} 
    \caption{
    \textbf{Principle of the Kondo cloud investigation in a dot-cavity device.}
    (a) The gate voltages $V_{\rm DL}$, $V_{\rm P}$, $V_{\rm UL}$, $V_{\rm UR}$ and $V_{\rm DR}$ control the QD potential and tunnel barriers, while $V_{\rm mirror}$ and $V_{\rm cav}$ control the cavity. Other gates are kept at zero voltage. One of the QD reservoirs is connected to the RF measurement setup, while the reservoir on the right is a grounded FP cavity. An electron microscope image of the device is shown in SM Fig.~S17.  
    (b) Simplified schematic showing the three regions of interest and their energy diagrams. $\Gamma_{\rm L}$ and $\Gamma_{\rm R}$ represent the tunneling rates between the QD and the reservoirs, $U$ the QD charging energy, $T_{\rm K}$ the Kondo temperature and $\Delta$ the cavity level spacing. The DOS in the FP cavity is shown for two different values of $V_{\rm cav}$ corresponding to large (max) and small (min) tunneling rate $\Gamma_{\rm R}$.
    (c) Stability diagram obtained by measuring the differential conductance as a function of gate and bias voltages (other parameters are given in SM Table~S1). Coulomb blockade diamonds are highlighted by dashed lines. The arrow indicates the Kondo ridge around zero bias.
    (d) Kondo peak in the center of the Kondo valley indicated in (c), plotted as a function of the cavity gate voltage which generates interference fringes. 
    (e) Kondo peak along the diagonal dashed lines in (d) after rescaling the bias voltage to extract $T_{\rm K}$ from the width of the conductance peak.
    }
    \label{fig:fig1}
\end{figure*}

%--------------------------------------------------------------
%\section{Experimental setup}
%--------------------------------------------------------------

\textit{Experimental setup} --- 
The Kondo QD is defined by electrostatic gates fabricated on the surface of a GaAs/AlGaAs heterostructure as shown in Fig.~\ref{fig:fig1}a. 
The two-dimensional electron gas (2DEG) located 100~nm below the surface has a $2.5\times10^{11}$~cm$^{-2}$ electron density and a $2.5\times10^{6}$~cm$^2$V$^{-1}$s$^{-1}$ electron mobility. 
The QD can be coupled to three electron reservoirs via the tunnel barriers formed by three quantum point contacts (QPCs). 
In this work, the QD is operated in a two-terminal configuration with the lower QPC always pinched off by means of a large negative voltage $V_{\rm DL}$. 
The common gate shared by the two remaining QPCs is set to a constant voltage $V_{\rm UR}$. 
The three gate voltages $V_{\rm DR}$, $V_{\rm P}$ and $V_{\rm UL}$ are used to control both the charge in the QD and the tunneling rates to the two reservoirs, denoted $\Gamma_{\rm L}$ and $\Gamma_{\rm R}$ for the left (L) and right (R) barriers, respectively.

The reservoir on the right of the QD is turned into a 3.2~$\mu$m-long FP cavity by applying a fixed voltage $V_{\rm mirror}=-800$~mV to the rightmost pair of gates, which fully depletes the 2DEG underneath and creates a hard-wall barrier.
Instead of varying the length of the cavity with $V_{\rm mirror}$, the FP interference is tuned by varying the electron wavelength in the 200 nm-long region below the gate located in the middle of the FP cavity, using a small negative voltage $V_{\rm cav}$ (no depletion). 
This results in a modulated density of states (DOS) at the Fermi level (as sketched in Fig.~\ref{fig:fig1}b), affecting the tunneling rate $\Gamma_{\rm R}$ between the QD and the right reservoir.
Note that electron transport in the cavity is ballistic, since the mean free path is about 20~$\mu$m in the 2DEG.

The conductance through the QD is measured using source-coupled radio-frequency (RF) reflectometry \cite{vigneau_probing_2023}. 
A 370~MHz sinusoidal wave is sent through a 50~$\Omega$ line to a resonant circuit connected to the ohmic contact of the left reservoir (see Fig.~\ref{fig:fig1}a). 
The amplitude (and phase) of the RF wave transmitted to the output port provides the admittance of the device.
In all graphs, the RF signal has been converted into differential conductance $G$, as explained in Supplemental Material (SM) section~1 \cite{SM}.
A DC source-drain bias $V_{\rm SD}$ is applied to the same contact using a bias tee.

\begin{figure*}
    \centering
    \includegraphics[width=\textwidth]{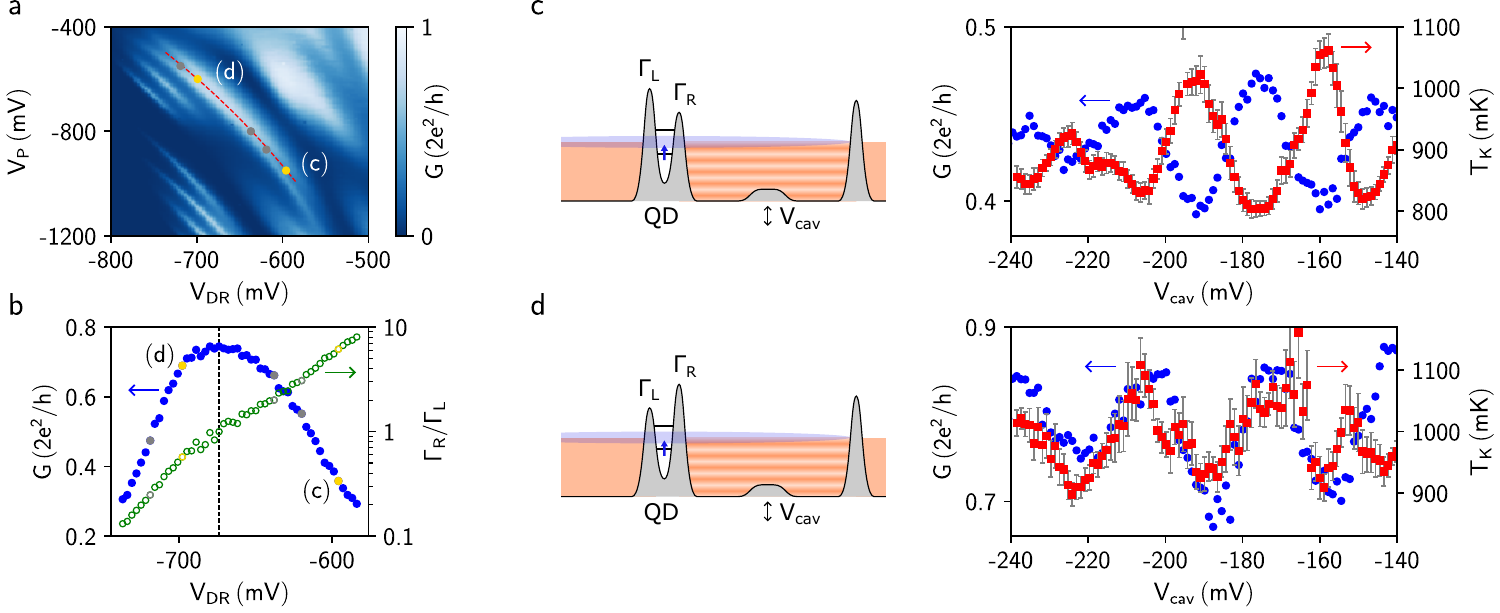}
    \caption{
    \textbf{Kondo temperature oscillations for opposite barrier asymmetries.} 
    (a) Zero-bias conductance map of the QD as a function of two gate voltages tuning both the charge number and the relative coupling to the two reservoirs. The white stripes correspond to Coulomb blockade peaks (the stability diagram in absence of cavity is shown in SM Fig.~S9b). The red dashed line indicates the Kondo valley where the investigation is performed. Along this line, from left to right, the barrier asymmetry changes from a smaller to a larger coupling to the cavity.
    (b) Evolution of the zero-bias conductance (blue dots) along the dashed line in (a). The tunneling rate ratio (green dots) is equal to 1 when the conductance is maximum (dashed line).
    (c,d) Kondo temperature (blue dots) and zero-bias conductance (red dots) as a function of the gate voltage tuning the interference in the cavity. These two quantities oscillate out of phase in (c) and in phase in (d) due to opposite barrier asymmetries, as represented in the schematics.
    }
    \label{fig:fig2}
\end{figure*}

%--------------------------------------------------------------
%\section{Kondo temperature measurements}
%--------------------------------------------------------------

\textit{Kondo temperature measurements} --- 
The stability diagram of the QD is shown in Fig.~\ref{fig:fig1}c using $V_{\rm DR}$ as plunger gate voltage to tune the QD charge state.
The Coulomb blockade diamond located around $V_{\rm DR}=-700$~mV exhibits the Kondo effect, as revealed by the presence of a horizontal ridge of enhanced conductance at zero source-drain bias. 
Additional horizontal lines are visible throughout the stability diagram, originating from FP interference in the cavity \cite{rossler_transport_2015,thimm_kondo_1999,dias_silva_conductance_2017}. 

These lines denote the alignment of a cavity mode with the Fermi energy of the source reservoir, either in the standard sequential tunneling regime along the diamond edges, or within the Coulomb diamonds due to strong elastic cotunneling \cite{franceschi_cotunneling_2001}.
The measured level spacing $\Delta \sim 150~\mu$eV is consistent with the relation $\Delta=hv_{\rm F}/2L=140~\mu$eV for a cavity length $L=3.2~\mu$m and a Fermi velocity $v_{\rm F}=2.2\times10^5$~m/s. 

The impact of tuning the FP interference with $V_{\rm cav}$ is shown in Fig.~\ref{fig:fig1}d at fixed $V_{\rm DR}$ in the middle of the Kondo valley.
Increasing $V_{\rm cav}$ shifts the ladder of FP resonances down in energy, such that the maxima of elastic cotunneling draw diagonal lines, while the Kondo peak remains at zero bias. 
Whenever a FP resonance crosses zero bias, it increases the DOS at the Fermi level, with an expected increase of $T_{\rm K}$ and a broadening of the Kondo peak.

To quantitatively analyze the changes in the Kondo effect caused by the FP interference, we need a reliable method to extract $T_{\rm K}$. 
This characteristic energy scale is usually determined via phenomenological scaling laws, either from the temperature dependence of the linear conductance \cite{goldhaber-gordon_kondo_1998}, or from the width of the Kondo peak in the non-linear differential conductance \cite{kretinin_universal_2012}. 
To avoid thermal broadening of the interfering electrons, and contrary to \cite{borzenets_observation_2020,tu_electrical_2024}, we employ here the second method based on the bias dependence, as illustrated in Fig.~\ref{fig:fig1}e. 

Since the interference fringes are tilted, the width of the Kondo peak is measured from line cuts parallel to the fringes (colored dashed lines in Fig.~\ref{fig:fig1}d). 
The scaling law employed to measure $T_{\rm K}$ is explained in details in SM section~2.1.
Basically, $T_{\rm K}$ is proportional to the width of the conductance peak, with a typical value around 1~K. 

The bias dependence has been checked to provide similar values as the temperature dependence in absence of cavity (see SM section~2.3), but is more appropriate in the present case where the total tunneling rate $\Gamma\sim400~\mu$eV is comparable to charging energy $U\sim500~\mu$eV (see SM section~2.4). 
The resulting charge fluctuations give indeed a parasitic contribution to the conductance which increases significantly with temperature, hampering a precise determination of $T_{\rm K}$ through temperature dependence (see SM section~2.2).

In addition, the bias dependence appears more reliable to extract $T_{\rm K}$ in situations where the cavity mode spacing $\Delta$ is comparable to $T_{\rm K}$.
The scaling analysis versus temperature indeed requires measuring the conductance up to temperatures of the order of $T_{\rm K}$, which induces thermal broadening of the conductance oscillations and affects the determination of the Kondo temperature (see SM section~4).
This effect is particularly significant when the cavity is longer than the Kondo cloud (but still of the same order), which is the case here since $L=3.2~\mu$m and $\xi_{\rm K}=1.7~\mu$m.

\begin{figure*}
    \centering
    \includegraphics[width=\textwidth]{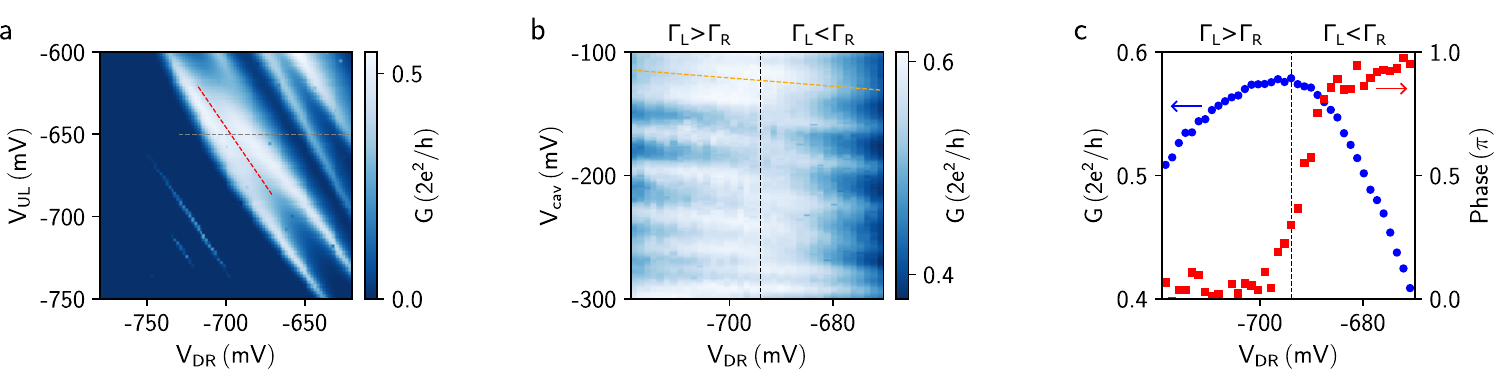}
    \caption{
    \textbf{Sign inversion of conductance oscillations for opposite barrier asymmetry.} 
    (a) Zero-bias conductance map of the QD as a function of two gate voltages tuning both the charge number and the relative coupling to the two reservoirs. The stability diagram along the grey dashed line is shown in Fig.~\ref{fig:fig1}c.
    (b) Evolution of the zero-bias conductance oscillations (tuned by $V_{\rm cav}$) as the barrier asymmetry is varied along the red dashed line in (a).
    (c) Blue dots: zero-bias conductance averaged over the oscillations shown in (b). Red squares: phase of the conductance oscillations shown in (b) after subtraction of the slope indicated by the orange dashed line. A phase shift by $\pi$ is observed when the conductance is maximum for symmetric barriers (black dashed line). 
    }
    \label{fig:fig3}
\end{figure*}

%--------------------------------------------------------------
%\section{Influence of barrier asymmetry}
%--------------------------------------------------------------

\textit{Influence of barrier asymmetry} --- 
We now turn to the specific topic of this work, which is to explore the combined influence of FP interference and barrier asymmetry on the Kondo effect.
To properly adjust the tunneling rates of the two barriers, a charge stability diagram is recorded as a function of the gate voltages $V_{\rm P}$ and $V_{\rm DR}$, controlling mainly the left and right barriers, respectively. 
The red dashed line in Fig.~\ref{fig:fig2}a indicates a Kondo valley, bounded by the Coulomb blockade peaks of the same spin-degenerate orbital state. 
In this Kondo valley, the conductance is weak for strong barrier asymmetry at the extremities of the line, and reaches almost $2e^2/h$ for symmetric barriers near the middle of the line, as shown in Fig.~\ref{fig:fig2}b.

Since the linear conductance in the Kondo regime is a resonant tunneling process through a many-body state at the Fermi level, its  value, in the low-temperature limit ($T \ll T_{\rm K}$), follows the general form \cite{pustilnik_kondo_2004}
\begin{equation}
    G \simeq \frac{2e^2}{h} \, \frac{4\Gamma_{\rm L}\Gamma_{\rm R}}{(\Gamma_{\rm L}+\Gamma_{\rm R})^2} \, ,
    \label{eq:conductance} 
\end{equation}
which strongly depends on the barrier asymmetry and reaches a maximum for symmetric barriers \cite{vanderwiel_kondo_2000}. 
This relation is used to extract the ratio $\Gamma_{\rm R}/\Gamma_{\rm L}$ plotted in Fig.~\ref{fig:fig2}b, assuming that $V_{\rm DR}$ preferentially tunes the tunneling rate $\Gamma_{\rm R}$ (see SM section~3). 

For selected positions along this line (dots in Fig.~\ref{fig:fig2}a), the Kondo peak is recorded versus cavity gate voltage (as in Fig.~\ref{fig:fig1}d) to extract the zero-bias conductance and the Kondo temperature as a function of the interference in the cavity.
The result is shown in Fig.~\ref{fig:fig2}c and \ref{fig:fig2}d for $\Gamma_{\rm R}/\Gamma_{\rm L} \sim 6$ and 0.6, respectively, and in SM Fig.~S16 for other values.
For all barrier asymmetries, $T_{\rm K}$ is found to oscillate, as a consequence of DOS modulations at the Fermi level in the cavity, which modulate the tunneling rate $\Gamma_{\rm R}$. 
This result demonstrates that the Kondo cloud spans the entire cavity, since the electrons of the cloud need to reach the mirror gate to make interference. 
It confirms the result of Ref.~\cite{borzenets_observation_2020} reporting a Kondo cloud extension of a few microns.

When comparing the two opposite barrier configurations, the most striking difference is that $T_{\rm K}$ oscillations and conductance oscillations are out-of-phase in Fig.~\ref{fig:fig2}c whereas they are in phase in Fig.~\ref{fig:fig2}d.
This finding can be understood by analyzing the specific dependence of each quantity to the tunneling rate $\Gamma_{\rm R}$, which is the parameter affected by the interference in the cavity.

In the Anderson model of a single-level impurity, the Kondo temperature in the center of the Kondo valley is given by \cite{haldane_atomic_1978} 
\begin{equation}
    T_{\rm K} = \frac{\sqrt{(\Gamma_{\rm L}+\Gamma_{\rm R})\,U}}{2}
    \,\exp\!{\left[-\frac{\pi\,U}{4\,(\Gamma_{\rm L}+\Gamma_{\rm R})}\right]} \, .
    \label{eq:temperature}
\end{equation}
This expression is maximum when $\Gamma_{\rm R}$ is maximum, which happens for constructive interference in the cavity for electrons at the Fermi level.
As opposed to the conductance given by Eq.~\eqref{eq:conductance}, $T_{\rm K}$ given by Eq.~\eqref{eq:temperature} does not depend on the barrier asymmetry, since only the sum of the tunneling rates enters the expression. 
$T_{\rm K}$ is therefore maximum for constructive interference both in Fig.~\ref{fig:fig2}c and \ref{fig:fig2}d.

On the other hand, the barrier asymmetry plays an important role in the sign of the conductance oscillations.
For $\Gamma_{\rm R}>\Gamma_{\rm L}$, constructive interference (larger $\Gamma_{\rm R}$) increases the barrier asymmetry, which reduces the transmission of the resonant tunneling process. 
Conversely, for $\Gamma_{\rm R}<\Gamma_{\rm L}$, constructive interference  (larger $\Gamma_{\rm R}$) reduces the barrier asymmetry, which increases the transmission. 
The conductance oscillations therefore have opposite sign in the two configurations.
They are out of phase (resp. in phase) with $T_{\rm K}$ oscillations in Fig.~\ref{fig:fig2}c (resp.~\ref{fig:fig2}d).

This sign inversion is examined in more detail in Fig.~\ref{fig:fig3} (for a slightly different configuration of gate voltages).
The charge stability diagram in Fig.~\ref{fig:fig3}a is recorded as a function of the gate voltages $V_{\rm UL}$ and $V_{\rm DR}$, controlling the left and right barriers, respectively. 
Along the red dashed line, the barrier asymmetry is continuously varied while keeping the QD in the center of the Kondo valley.
Figure~\ref{fig:fig3}b shows the evolution of the conductance oscillations (induced by the cavity) along this path, revealing a clear phase shift in the middle of the map.
Using a Fourier analysis to track the phase of the oscillations, and after removing the slope caused by the gradual change of cavity length with $V_{\rm DR}$ (orange dashed line), an abrupt phase shift by $\pi$ is observed in Fig.~\ref{fig:fig3}c. 
This sign inversion of the conductance oscillations coincides with the maximum of conductance, at which the two barriers are symmetric, with a QD equally coupled to both reservoirs. 
At this particular point (vertical dashed line), the highly transmitting QD is a poorly reflecting mirror for the FP cavity and the interference fringes almost disappear.

%--------------------------------------------------------------
%\section{Numerical modeling}
%--------------------------------------------------------------

\begin{figure}
    \centering
    \includegraphics[width=0.5\textwidth]{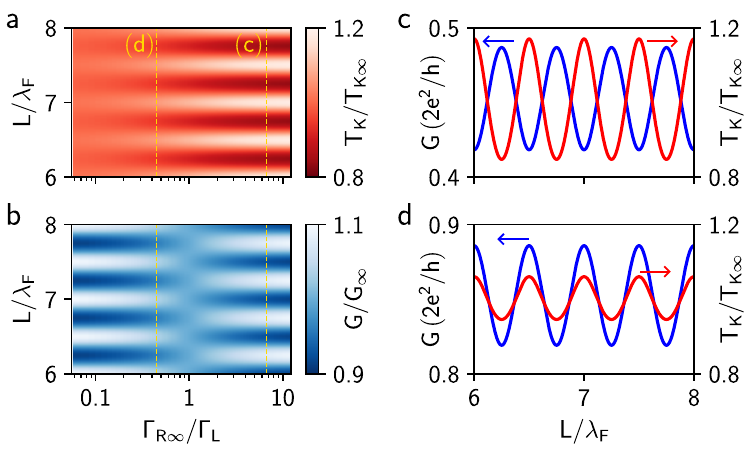}
    \caption{
    \textbf{Numerical modeling of $G$ and $T_{\rm K}$ oscillations.}
    (a) Kondo temperature normalized to its value in absence of cavity, as a function of tunneling rate ratio and cavity length normalized to the Fermi wavelength.
    (b) Conductance of the dot-cavity system normalized to its value in absence of cavity, as a function of the same parameters.
    (c,d) Conductance (blue line) and Kondo temperature (red line) versus cavity length, for barrier asymmetries indicated by dashed lines in (a,b).
    }
    \label{fig:fig4}
\end{figure}

\textit{Numerical modeling} --- 
The experimental observations can be reproduced numerically by simulating the Kondo conductance with Eq.~\eqref{eq:conductance} and the Kondo temperature with Eq.~\eqref{eq:temperature}, in which the tunneling rate to the cavity is expressed as
\begin{equation}
    \Gamma_{\rm R} = \Gamma_{\rm R\infty} \, \big(1+\alpha\cos(2k_{\rm F}L)\big) \, ,
    \label{eq:gamma}
\end{equation}
where $\Gamma_{\rm R\infty}$ is the tunneling rate in absence of cavity, $k_{\rm F}$ the Fermi wave vector, $L$ the cavity length, and $\alpha$ the contrast of the interference controlled by the reflection coefficient of the mirror located to the far right of the cavity (see SM section~4.3 and Ref.~\cite{park_how_2013}). 

Figures~\ref{fig:fig4}a and \ref{fig:fig4}b display the resulting $T_{\rm K}$ and conductance oscillations as a function of the barrier asymmetry parameter $\Gamma_{\rm R\infty}/\Gamma_{\rm L}$.  
As this ratio increases, the visibility of the $T_{\rm K}$ oscillations gradually increases, because the Kondo effect becomes increasingly dominated by the coupling $\Gamma_{\rm R}$ to the cavity (see SM section~6).
On the other hand, the conductance exhibits a phase shift by $\pi$ when the barrier asymmetry is reversed, as evidenced experimentally in Fig.~\ref{fig:fig3}b.

The relative sign of $T_{\rm K}$ and conductance oscillations is visualized in Fig.~\ref{fig:fig4}c and \ref{fig:fig4}d for opposite barrier asymmetries. 
The graphs display the same out-of-phase and in-phase oscillations as in the experiment shown in Fig.~\ref{fig:fig2}c and \ref{fig:fig2}d for the same asymmetry parameters.

%--------------------------------------------------------------
%\section{Conclusions}
%--------------------------------------------------------------

\textit{Conclusions} --- 
In this work, we investigated the spatial extension of the Kondo screening cloud by coupling a 3.2~$\mu$m-long cavity to a Kondo QD. 
We observed Kondo temperature oscillations while tuning the interference in the cavity. 
These oscillations, with maxima occurring for constructive interference, provide direct evidence that the Kondo cloud extends over micron-scale distances, consistent with a theoretical Kondo length of 1.7~$\mu$m.

Furthermore, the oscillations of the zero-bias conductance are shown to be either in phase or out of phase with those of the Kondo temperature, depending on the coupling asymmetry of the QD with the two reservoirs. 
This effect highlights the critical role of barrier asymmetry for phase-coherent transport in the Kondo regime.

Our results provide new experimental information on the properties of the Kondo cloud, paving the way for future investigations of QD arrays with overlapping Kondo clouds, as tunable platforms for the simulation of quantum materials containing multiple magnetic impurities, or involving multi-channel Kondo effect \cite{iftikhar_superballistic_2018, pouse_quantum_2023, shim_hierarchical_2023}.
The geometry of a Kondo dot coupled to a FP cavity should also enable measurements of the Kondo reflection phase, thereby complementing previous investigations of the Kondo transmission phase using Aharonov-Bohm interferometers \cite{ji_phase_2000,takada_transmission_2014}.

%--------------------------------------------------------------
\section{Acknowledgments}
%--------------------------------------------------------------

We thank C. Bäuerle and S. Florens for the illuminating discussions, F. Gay and the LATEQS technical staff for their assistance during the measurements. This work received financial support from the Agence Nationale de la Recherche under the contract ANR-18-CE30-0027 ``KONEX''. B.H. (senior research associate) acknowledges support from the F.R.S-FNRS as well as from the ARC project DREAMS (21/26.116).

\section{Data availability 
}
%--------------------------------------------------------------
The data that support the findings of this article are
openly available at  \url{https://doi.org/10.5281/zenodo.17475869}

%--------------------------------------------------------------

%--------------------------------------------------------------
\end{document}

\vfill
\begin{center}
    \large
    \textbf{SUPPLEMENTAL MATERIAL} \\
    \vspace{5mm}
    for \\
    \vspace{5mm}
    \textbf{Kondo cloud conductance in cavity-coupled quantum dots \\ with asymmetric barriers} \\
    \vspace{5mm}
    by Fossion \textit{et al.}
    \vspace{5mm}
\end{center}
\vfill
\tableofcontents
\thispagestyle{empty}
\vfill
\pagebreak

%--------------------------------------------------------------
\noindent\hrulefill\vspace{-5mm}
\section{Radio-frequency measurements}
\vspace{-5mm}\noindent\hrulefill
%--------------------------------------------------------------

%--------------------------------------------------------------
\subsection{Experimental setup}
%--------------------------------------------------------------

\begin{figure}[!h]
    \centering
    \includegraphics[width=12cm]{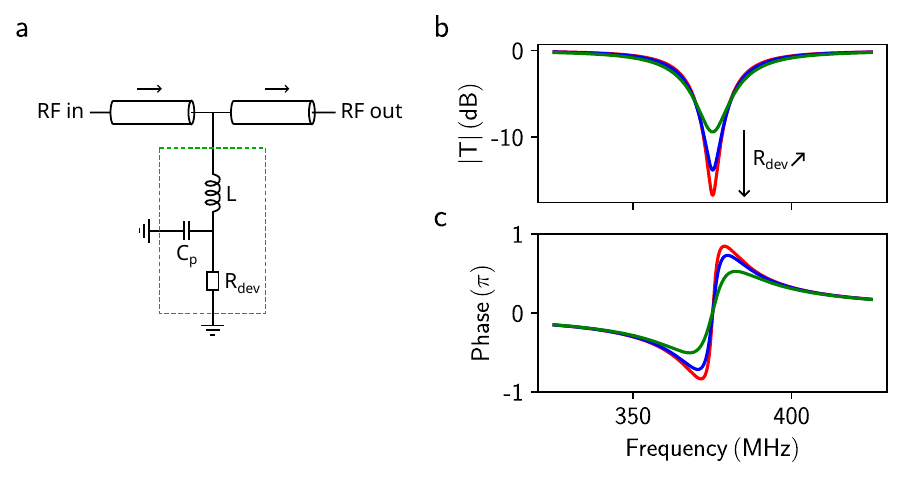}
    \caption{
    \textbf{Setup for RF measurements.} 
    (a) Quantum device embedded in a matching circuit and connected to the transmission line. The inductance $L$ is a surface-mount element while the capacitance $C_{\rm p}$ is due to parasitic capacitive couplings to ground on the mounting board and in the device. The bias tee is not shown here.
    (b,c) Magnitude and phase of the transmission coefficient for different $R_{\sf dev}$: Effect of variation of $R_{\sf dev}$ on the transmission amplitude (a) and phase (b). The curves are plotted for $L=240$~nH and $C_{\rm p}=0.75$~pF.
    }
    \label{RF_setup} 
\end{figure}

Radio-frequency (RF) measurements provide a fast and sensitive method for measuring transport properties of quantum dots. 
The key parameter to get a large RF response is the matching between the impedance $Z_0=50~\Omega$ of the RF line and the impedance of the circuit containing the device to be measured. 
Since the resistance of a quantum dot is larger than 10~k$\Omega$, a matching circuit is needed.
For this purpose, a LC resonator is inserted between the RF line and an ohmic contact of the device (called source), as shown in Fig.~\ref{RF_setup}a, and another ohmic contact of the device (called drain) is connected to ground.
The resonator is composed of a surface-mount inductor $L\sim240$~nH and a parasitic capacitance $C_{\rm p}\sim0.75$~pF in parallel with the device resistance $R_{\rm dev}$.

Close to the resonance frequency $f_{\rm res}=1/2\pi\sqrt{LC_{\rm p}}$, the circuit can be simplified to a series RLC circuit of effective resistance $R_{\rm eff}=L/C_{\rm p}R_{\rm dev}$.
The resonant circuit is absorbing part of the RF signal traveling through the line, which creates a dip in the transmission amplitude.
Variations in the device resistance change the depth of the dip in the RF amplitude and the slope in the RF phase, as shown in Fig.~\ref{RF_setup}b,c.
Even slight modifications to the device resistance result in substantial changes in the RF signal.
The best sensitivity in amplitude measurements is obtained at the resonance frequency of the resonator.
For the typical values $L=240$~nH, $C_{\rm p}=0.75$~pF and $R_{\rm dev}=26$~k$\Omega$, the effective circuit resistance is $R_{\rm eff}\sim Z_0/4$, providing a good matching with the RF line and thus a large RF response.

The RF wave applied at room temperature is attenuated by a total of 60~dB along the fridge stages, and reaches the device at low temperature with an amplitude of about 6~$\mu$V, enabling linear conductance sensing at the base temperature of 50~mK. 
The RF wave transmitted by the resonant circuit is then amplified by 72~dB and measured with an ultra-high-frequency lock-in amplifier.

%--------------------------------------------------------------
\subsection{Examples of RF signals for QPC and QD}
%--------------------------------------------------------------

\begin{figure}[!h]
    \centering
    \includegraphics[width=14cm]{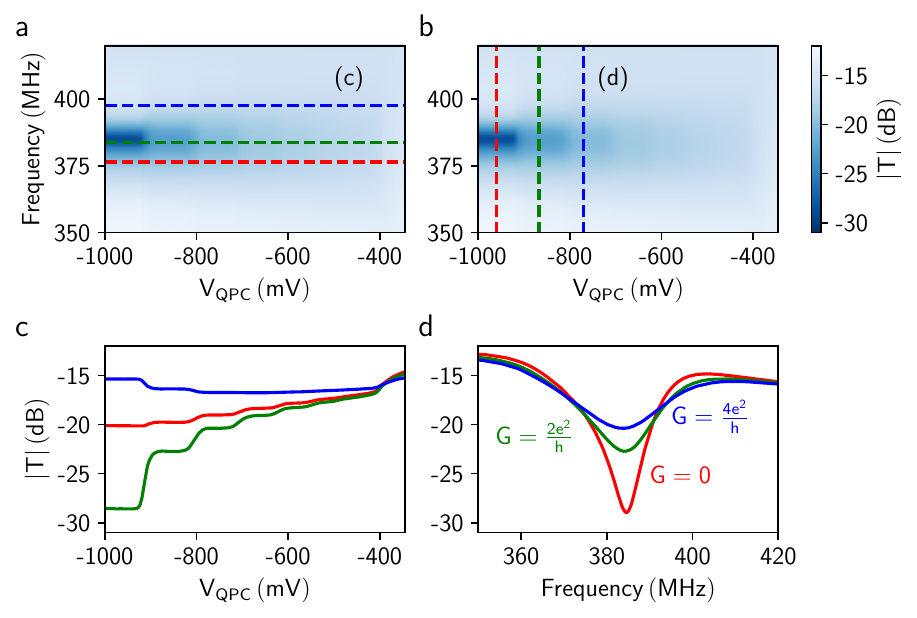}
    \caption{
    \textbf{RF measurement of a QPC.}
    (a,b) Transmission $|T|$ as a function of the frequency and the QPC gate voltage. 
    (c) Cut along the dashed lines from (a). At the resonance frequency, the green curve, corresponding to a frequency of 383~MHz, shows well-defined plateaus. Slightly changing the frequency reduces the visibility of the plateaus, as shown by the red curve for a frequency of 376~MHz, or can even change the expected behavior, with the curve going up while the QPC is becoming increasingly closed, as shown by the blue curve at a frequency of 397~MHz.
    (d) Cut along the dashed lines from figure (b). The depth of the resonance depends monotonously on the QPC conductance, with the largest variations obtained at the resonance frequency.
    }
    \label{Device2_Reflectoo} 
\end{figure}

Figure~\ref{Device2_Reflectoo} shows the amplitude of the transmitted RF signal for a QPC. 
At the resonance frequency (green curve in Fig.~\ref{Device2_Reflectoo}c), the transmission $|T|$ shows well-defined plateaus as expected for the quantized conductance of a QPC.
The system detects correctly any change in the device resistance when $V_{\rm QPC}$ is changed. 
However, slightly shifting the frequency compared to $f_{\rm res}$ reduces the visibility of the plateau (red curve in Fig.~\ref{Device2_Reflectoo}c) or even leads to the opposite of the expected behavior, where the curve starts to go up when the QPC is becoming increasingly closed (blue curve in Fig.~\ref{Device2_Reflectoo}c).
This behavior can be understood by looking at panel (d) showing an inversion of the response for highly detuned frequencies, as a result of an increased resonance width for large device conductance.

\begin{figure}[!h]
   \centering
   \includegraphics[width=14cm]{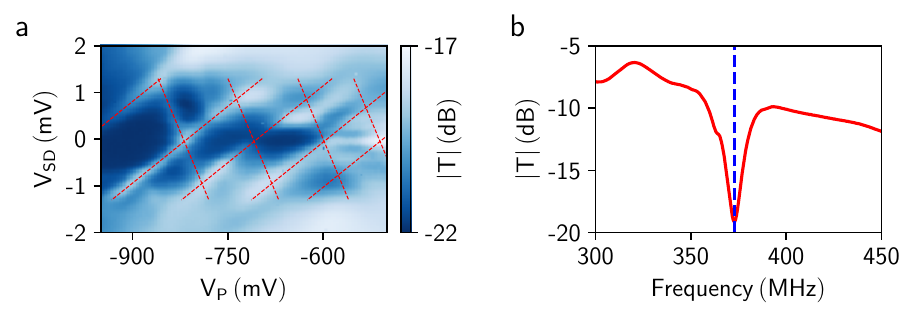}
   \caption{
   \textbf{RF measurement of a QD.}
   (a) Evolution of the RF transmission as a function of $V_{\rm SD}$ and $V_{\rm P}$ at a frequency of 372.8~MHz. It shows several Coulomb diamonds (red dashed lines) and a Kondo resonance at zero bias inside the diamond centered around $V_{\rm P}=-560$~mV.
   (b) Evolution of the RF transmission with frequency for $V_{\rm P}\sim-1000$~mV, showing a resonance at $f_{\rm res}=372.8$~MHz.
   }
   \label{ProbeQuantumDot}
\end{figure}

Figure~\ref{ProbeQuantumDot}a shows the stability diagram of a QD probed by RF reflectometry.
A succession of Coulomb blockade diamonds is visible (red dashed lines) with a Kondo resonance at zero bias in the diamond centered around $V_{\rm P}=-560$~mV. 
The measurement frequency was fixed at 372.8~MHz, corresponding to the resonant frequency of the circuit as shown in Fig.~\ref{ProbeQuantumDot}b. 

%--------------------------------------------------------------
\subsection{Conversion of RF signal into differential conductance}
%--------------------------------------------------------------

\begin{figure}[!h]
    \centering
    \includegraphics[width=14cm]{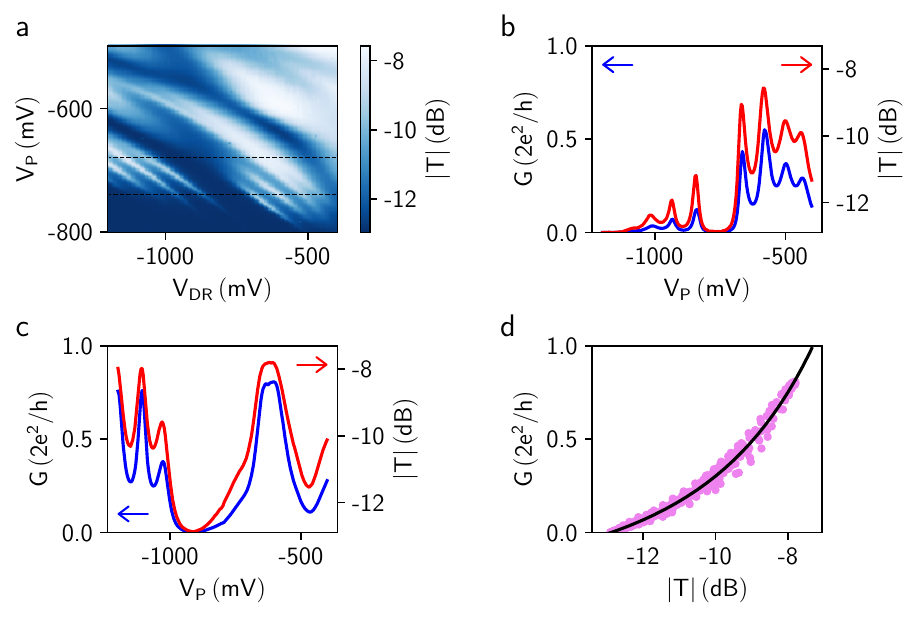}
    \caption{
    \textbf{Conversion from RF signal to differential conductance.} 
    (a) Evolution of $|T|$ as a function of $V_{\rm DR}$ and $V_{\rm P}$ at a frequency of $372.7$~MHz, highlighting Coulomb peaks. 
    (b,c) Cuts of the transmission map (red curves) along the dashed lines in (a) at $V_{\rm P}=-740$~mV and $-680$~mV, respectively. In addition, the differential conductance was measured with $V_{\rm AC}=10~\mu$V at 77~Hz. A resistance of 20~k$\Omega$ is removed to take into account the ohmic contacts and the fridge line resistances. 
    (d) Differential conductance as a function of $|T|$. The gray dots are the data from both (b) and (c), and the black line is the extracted conversion curve.
    }
    \label{Device2_Reflecto}
\end{figure}

A drawback of RF measurements is that the value of the conductance and the related physical quantities (barrier asymmetry, Kondo temperature, etc...) are not directly accessible.
Therefore, converting the RF signal to a differential conductance in units of $e^2/h$ is essential for quantitative analysis of the data. 
This conversion is demonstrated here for a quantum dot. 
In Fig.~\ref{Device2_Reflecto}a, the RF transmission $|T|$ is plotted against $V_{\rm P}$ and $V_{\rm DR}$. 
For two specific values of $V_{\rm P}$ (indicated by black dashed lines Fig.~\ref{Device2_Reflecto}a), the low-frequency differential conductance ${\rm d}I/{\rm d}V$ is measured simultaneously with the RF transmission by applying $V_{\rm AC}=10~\mu$V through a bias-tee and by recording the AC current with an amplifier connected to an opposite ohmic contact. 
The conductance is then corrected by a 20~k$\Omega$ series resistance to account for the resistances of the ohmic contacts and the fridge lines. 
This dual measurement is illustrated in Fig.~\ref{Device2_Reflecto}b and c, where the RF signal amplitude is shown in red and the differential conductance in blue. 
As anticipated, the two curves exhibit similar behavior. 
By plotting the differential conductance against its corresponding $|T|$ value (dots in Fig.~\ref{Device2_Reflecto}d), a conversion curve can be derived, allowing any RF measurement to be converted into differential conductance. 
Each time the frequency, RF power, or sample is changed, a new calibration is necessary. 
In this specific example, the calibration is valid for the sample used in the data presented in this article at $f=372.7$~MHz. 

\newpage

\clearpage
%--------------------------------------------------------------
\noindent\hrulefill\vspace{-5mm}
\section{Extraction of the Kondo temperature from the bias dependence}
\vspace{-5mm}\noindent\hrulefill
%--------------------------------------------------------------

%--------------------------------------------------------------
\subsection{Fitting procedure of the Kondo resonance}
%--------------------------------------------------------------

A characteristic signature of the Kondo effect is the conductance peak observed around zero-bias in the Kondo valley of the stability diagram, as shown in Fig.~1c in the main text.
The differential conductance is maximum at zero-bias and is progressively reduced when the source-drain bias increases.
This reduction is caused by the inelastic scattering that reduces the coherence of the Kondo process, which gives an intrinsic width of the Kondo spectral function, related to the Kondo temperature $T_{\rm K}$.
The width of the conductance peak in the non-linear conductance is also related to the Kondo temperature $T_{\rm K}$, thus providing an experimental method to extract it from the bias dependence of the conductance.

In numerous experiments, the full width at half maximum (FWHM) was assumed to be equal to $2k_{\rm B}T_{\rm K}$ \cite{goldhaber_kondo_1998,cronenwett_tunable_1998,vandewiel_unitary_2000,sasaki_integer_2000,jespersen_kondo_2006}, but Kretinin \textit{et al.} \cite{kretinin_universal_2012} pointed out that this correspondence tends to overestimate $T_{\rm K}$ compared to the values obtained from the temperature dependence. 
They suggested instead to use $2\sqrt{\pi}k_{\rm {\rm B}}T_{\rm K}$ for the FWHM, based on the phenomenological equation
\begin{equation}
    G(V_{\rm SD}) = G_{\rm max} 
    \left[1+(2^{1/s_1}-1)\left(\frac{eV_{\rm SD}}{\sqrt{\pi}k_{\rm B}T_{\rm K}}\right)^2\;\right]^{-s_1}
    \label{Kretinin_Bias_Fit}
\end{equation}
which reproduces the results of NRG calculations at zero temperature with $s_1=0.32$. 

Unfortunately, the Kondo resonance in GaAs QDs usually does not have the line shape of Eq.~\eqref{Kretinin_Bias_Fit} and is surrounded by a significant cotunneling background.
In our work, we therefore used a specific procedure to extract $T_{\rm K}$ from the bias dependence of the conductance, as explained below.

\begin{figure}[!h]
    \centering
    \includegraphics[width=8cm]{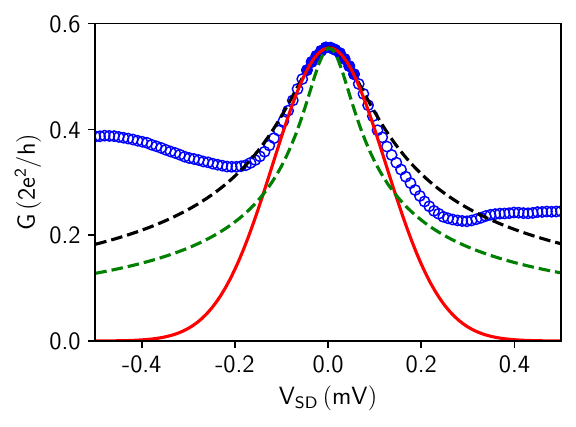}
    \caption{
    \textbf{Fitting procedure to analyze the Kondo resonance.}
    The blue symbols show the differential conductance measured as a function of source-drain bias in the Kondo regime. The top of the Kondo peak (filled symbols) is fitted using either the empirical equation~\eqref{Kretinin_Bias_Fit} (black dashed line) or a Gaussian function (red line). 
    }
    \label{Bias_fitting_discussion}
\end{figure}

We consider the measured Kondo resonance shown as symbols in Fig.~\ref{Bias_fitting_discussion}.
The black dashed line is the empirical equation~\eqref{Kretinin_Bias_Fit}.
Only the data points shown as filled dots are used for the fitting since the Kondo resonance is quickly broadened by the cotunneling background. 
This function correctly captures the behavior at low bias, but the measured conductance decreases faster than predicted by the model, leading to an overestimation of the Kondo temperature defined from the FWHM, far away from zero bias.
Using a Gaussian fitting curve represented by the red line, both the curvature at zero bias and the rapid decrease in conductance with bias are well captured. 
While both fitting functions correctly describe the behavior at low bias, the Gaussian fit captures better the rapid decrease of the conductance.

Since the Gaussian function has a smaller FWHM than the model, it would indicate a smaller Kondo temperature.
To check if the relation FWHM $=2\sqrt{\pi}k_{\rm {\rm B}}T_{\rm K}$ with the Gaussian fit could be a good definition of $T_{\rm K}$, we should compare the value $T_{\rm K,Bias}$ extracted from the bias dependence using this definition, and the value $T_{\rm K,Temp}$ extracted from the conventional temperature dependence.
In the next section, we show that these two values are very close, such that this definition is indeed correct.
For this reason, the fitting function used in our work to extract $T_{\rm K}$ is
\begin{equation}
    G(V_{\rm SD}) = G_{\rm max} 
    \,\exp\!\left[-\ln(2)\left(\frac{eV_{\rm SD}}{\sqrt{\pi}k_{\rm B}T_{\rm K}}\right)^2\;\right]
    \label{Gaussian_Bias_Fit}
\end{equation}

Finally, one can notice that the value of $T_{\rm K}$ obtained with this procedure corresponds to having the model curve from Eq.~\eqref{Kretinin_Bias_Fit} as represented by the green dashed line (with the same FWHM as the Gaussian curve) which is narrower at the top. 
Looking at the figure, the red curve appears as a compromise between the black curve (matching at low bias) and the green curve (matching at half maximum).

\clearpage
%--------------------------------------------------------------
\subsection{Extraction from the temperature dependence}
%--------------------------------------------------------------

The dependence on temperature of the zero-bias conductance in the Kondo regime is usually modeled by the phenomenological expression \cite{goldhaber-gordon_kondo_1998}
\begin{equation}
    G(T) = G_{\rm max} 
    \left[1+(2^{1/s_2}-1)\left(\frac{T}{T_{\rm K}}\right)^2\;\right]^{-s_2} 
    \label{Temperature_Fit}
\end{equation}
which reproduces the exact dependence from NRG calculations using $s_2=0.22$ for a spin 1/2.
This expression will be used below to extract $T_{\rm K,Temp}$ from the measured temperature dependence of the conductance.

\begin{figure}
    \centering
    \includegraphics[width=14cm]{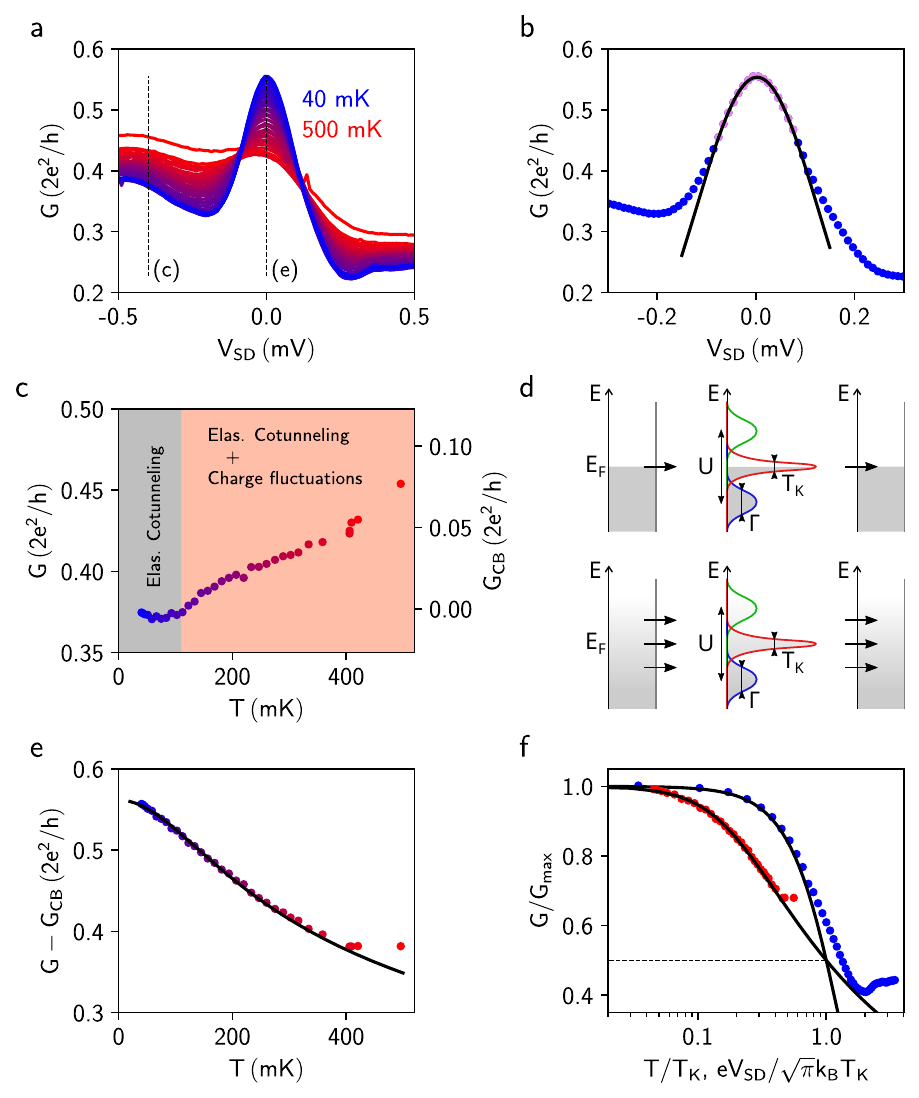}
    \caption{
    \textbf{Kondo temperature extracted from bias and from temperature dependencies.}
    (a) Differential conductance of a Kondo resonance for temperatures varying from 40~mK (blue) to 500~mK (red). 
    (b) Kondo resonance at $T=40$~mK with a Gaussian fit (black line) using Eq.~\eqref{Gaussian_Bias_Fit}, giving $T_{\rm K,Bias}=960$~mK.
    (c) Differential conductance as a function of temperature at $V_{\rm SD}=-0.4$~meV. The right axis indicates the contribution $G_{\rm CB}$ from the thermally activated conductance of the Coulomb blockade regime.
    (d) Energy diagram of electron tunneling through a QD at zero (top) and finite (bottom) temperatures in the regime $\Gamma\lesssim U$ with significant charge fluctuations.
    (e) Kondo contribution $G-G_{\rm CB}$ as a function of temperature with a fit (black line) using Eq.~\eqref{Temperature_Fit}, giving $T_{\rm K,Temp}=890$~mK. 
    (f) Normalized conductance as a function of the rescaled temperature $T/T_{\rm K}$ (red dots) and the rescaled bias $eV_{\rm SD}/\sqrt{\pi}k_{\rm B}T_{\rm K}$ (blue dots).
    }
    \label{TKBIAS_TKTEMP}
\end{figure}

Figure~\ref{TKBIAS_TKTEMP}a shows the Kondo resonance for different fridge temperatures from 40 to 500~mK.
The data at 40~mK are fitted using a Gaussian function as explained above and shown in Fig.~\ref{TKBIAS_TKTEMP}b.
This gives a Kondo temperature $T_{\rm K,Bias}=960$~mK, easily obtained from the bias dependence. 
On the other end, the extraction of $T_{\rm K,Temp}$ from the temperature dependence is more complicated, due to the presence of an additional contribution to the conductance at higher temperature, as explained below.

At very low temperature ($k_{\rm B}T \ll U-\Gamma$), the residual conductance in a Coulomb blockade diamond is caused by elastic cotunneling processes. 
At finite bias, outside the Kondo resonance, the spin-independent elastic cotunneling is independent of temperature \cite{cox_lowtemperature_1997,glazman_lowtemperature_2005,jiang_spin_2007}.
This is observed for temperatures below 100~mK in Fig.~\ref{TKBIAS_TKTEMP}c obtained at $V_{\rm SD}=-0.4$~meV with a constant non-zero conductance. 
For temperatures higher than 100~mK, the conductance progressively increases. 
The thermal energy in the reservoirs is sufficient to allow an extra electron to enter the dot by overcoming the charging energy $U$ (charge quantization is lifted when $k_{\rm B}T \sim U-\Gamma$). 
This extra conduction channel adds up to elastic cotunneling processes, with a contribution $G_{\rm CB}$ given by the right axis of Fig.~\ref{TKBIAS_TKTEMP}c, which corresponds to the temperature dependence of the Coulomb blockade conductance.

At zero bias, the usual cotunneling is replaced by the Kondo effect (enhanced cotunneling with spin-flip) with a strong temperature dependence on the scale $T_{\rm K}$. 
Since the thermally activated Coulomb blockade (transfer of several electrons) represents an uncorrelated channel for electron transport, this contribution adds up to the Kondo effect, as schematically shown in Fig.~\ref{TKBIAS_TKTEMP}d.
Consequently, to isolate the evolution of conductance due only to the Kondo effect, the additional contribution $G_{\rm CB}$ is removed from the measured conductance before fitting the data. 
The evolution of the corrected zero-bias conductance $G-G_{\rm CB}$ is plotted versus temperature in Fig.~\ref{TKBIAS_TKTEMP}e and fitted with the empirical equation~\eqref{Temperature_Fit}, providing $T_{\rm K,Temp}=890$~mK. 

This value is consistent with the value $T_{\rm K,Bias}=960$~mK obtained from the bias dependence using the procedure described in the previous section.
This consistency between the two methods validates the procedure of extracting $T_{\rm K}$ from the bias dependence using a Gaussian function and a FWHM $=2\sqrt{\pi}k_{\rm {\rm B}}T_{\rm K}$.

Finally, the two scaled dependencies, versus temperature and versus bias, are compared in Fig.~\ref{TKBIAS_TKTEMP}f. 
This graph highlights the difference in influence that these two parameters have on the Kondo conductance.

\clearpage
%--------------------------------------------------------------
\subsection{Comparison of the two methods for different gate configurations}
%--------------------------------------------------------------

\begin{figure}[!h]
    \centering
    \includegraphics[width=14cm]{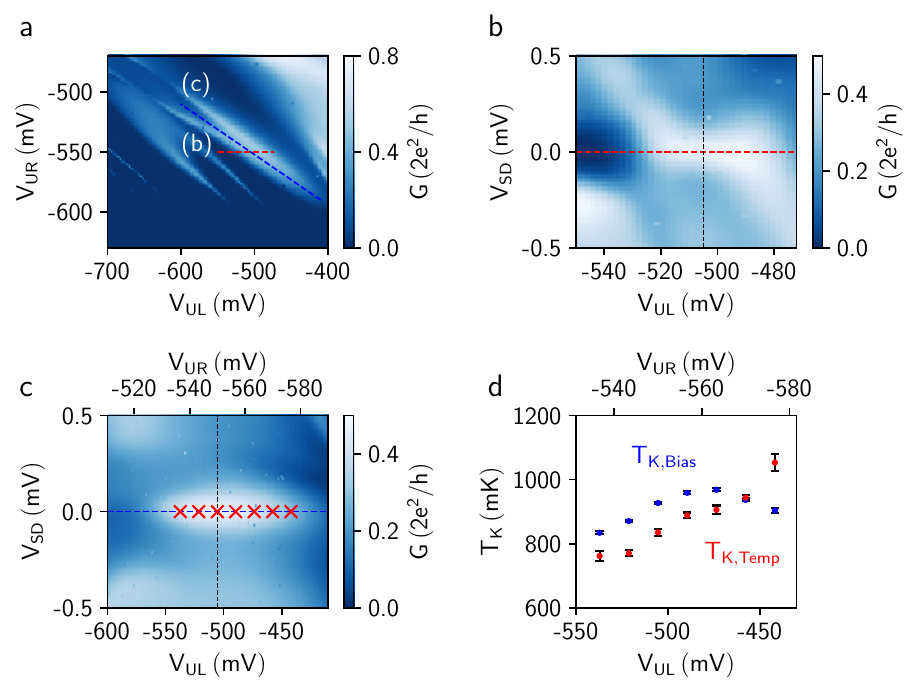}
    \caption{
    \textbf{Kondo temperatures from the two methods for different gate configurations.}
    (a) Differential conductance as a function of the gate voltages $V_{\rm UR}$ and $V_{\rm UL}$.
    (b) Energy stability diagram along the red dashed line in (a), showing a Kondo ridge at zero bias.
    (c) Evolution of the Kondo resonance along the blue dashed line in (a) in the middle of the Kondo valley. The intersection between the maps in (b) and (c) is indicated by the black dashed line.
    (d) Kondo temperatures measured for the gate configurations indicated by the red crosses in (c) and corresponding to different barrier asymmetries. The values $T_{\rm K,Temp}$ obtained by varying the temperature are shown in red and the values $T_{\rm K,Bias}$ obtained from the bias dependence are shown in blue.
    }
    \label{TK_Final}
\end{figure}

In order to check the robustness of the procedure used to extract $T_{\rm K}$ from the bias dependence, we compare the values $T_{\rm K,Bias}$ and $T_{\rm K,Temp}$ for different gate configurations.
Figure~\ref{TK_Final}a displays the differential conductance as a function of the gate voltages $V_{\rm UR}$ and $V_{\rm UL}$, with anti-diagonal stripes corresponding to Coulomb blockade peaks. 
The energy stability diagram along the red dashed lines is shown in Fig.~\ref{TK_Final}b, with a Kondo resonance at zero bias inside a Coulomb blockade diamond.  
The Kondo resonance along the blue dashed line is shown in Fig.~\ref{TK_Final}c, where the barrier asymmetry is continuously varied while keeping the QD in the middle of the Kondo valley.
The red crosses represent the pairs $(V_{\rm UL},V_{\rm UR})$ at which the temperature has been varied to extract $T_{\rm K,Temp}$, for a comparison with $T_{\rm K,Bias}$, as done in the previous section.
Note that the configuration $(V_{\rm UL},V_{\rm UR})=(-489,-557)$~mV corresponds to the analysis already shown in Fig.~\ref{TKBIAS_TKTEMP}.
The results are plotted in Fig.~\ref{TK_Final}d, showing a good match between the Kondo temperatures derived from bias and temperature dependencies.
The error bars represent the uncertainty of the fit.

\clearpage
%--------------------------------------------------------------
\subsection{Kondo temperatures along the Kondo ridge}
%--------------------------------------------------------------

\begin{figure}[!h]
    \centering
    \includegraphics[width=13.5cm]{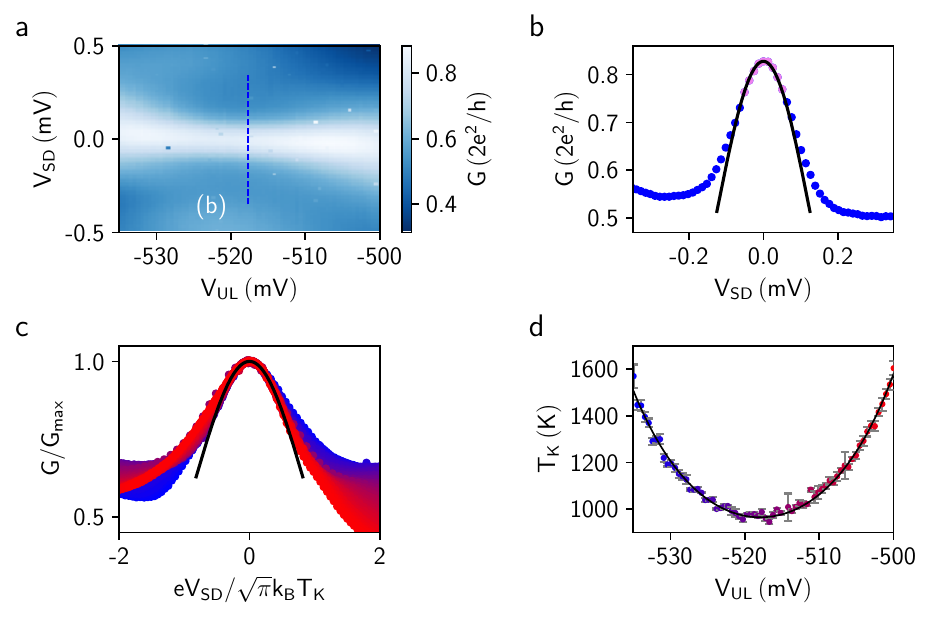}
    \caption{
    \textbf{Extraction of the Kondo temperature from the bias dependence in a Kondo valley.} 
    (a) Differential conductance as a function of $V_{\rm UL}$ and $V_{\rm SD}$ highlighting a Kondo ridge in a Coulomb blockade diamond. 
    (b) Gaussian function (in black) used to fit the Kondo resonance along the blue dashed line in (a). Only red points are considered for the fit as the background signal becomes rapidly significant. 
    (c) Universal dependence of the normalized conductance $G/G_{\rm max}$ as a function of the rescaled bias voltage, for different gate voltages $V_{\rm UL}$. 
    (d) Kondo temperature obtained from the scaling shown in (c) and plotted as a function of $V_{\rm UL}$. The data are fitted (black line) with the Haldane's formula~\eqref{Haldane_full}.
    }
    \label{bias_Dependence_Kondo_Sample3TG5}
\end{figure}

The extraction of $T_{\rm K}$ from the width of the Kondo resonance in a Kondo valley is performed in Fig.~\ref{bias_Dependence_Kondo_Sample3TG5}.
The Kondo temperature is extracted at different voltages $V_{\rm UL}$, playing the role of the plunger gate, along the Kondo ridge shown in Fig.~\ref{bias_Dependence_Kondo_Sample3TG5}a.
An example of fit is shown in Fig.~\ref{bias_Dependence_Kondo_Sample3TG5}b for $V_{\rm UL}=-518$~mV using the Gaussian function. 
Performing the same fit for all values of $V_{\rm UL}$ in the Kondo valley yields the scaling shown in Fig.~\ref{bias_Dependence_Kondo_Sample3TG5}c. 
This figure shows the normalized conductance $G/G_{\rm max}$ as a function of the scaled bias $eV_{\rm SD}/\sqrt{\pi}k_{\rm B}T_{\rm K}$.
After scaling, all the curves follow the same universal behavior. 
The values of $T_{\rm K}$ as a function of $V_{\rm UL}$ are plotted in  Fig.~\ref{bias_Dependence_Kondo_Sample3TG5}d.
Interestingly, this gate dependence can be fitted (black curve) with the Haldane's formula \cite{haldane_atomic_1978} 
\begin{equation}
    T_{\rm K} = \frac{\sqrt{\Gamma U}}{2} 
    \,\exp\!\left[\frac{\pi\epsilon_{\rm d}(\epsilon_{\rm d}+U)}{\Gamma U}\right]
\label{Haldane_full}
\end{equation}
with $\epsilon_{\rm d}=-\alpha_{\rm UL}\,(V_{\rm UL}-V_{\rm UL,0})$ the energy of the quantum state relative to the Fermi energy (usually negative), $\Gamma$ the total tunneling rate (defined as the FWHM of the quantum level) and $U$ the charging energy.
Using $\alpha_{\rm UL}=0.011$ and $U=450~\mu$eV deduced from an energy stability diagram, this fitting provides the value $\Gamma=370~\mu$eV.

\clearpage
%--------------------------------------------------------------
\noindent\hrulefill\vspace{-5mm}
\section{Extraction of the tunneling rates from the Kondo effect}
\vspace{-5mm}\noindent\hrulefill
%--------------------------------------------------------------

%--------------------------------------------------------------
\subsection{Extraction of the tunneling rate ratio from the Kondo conductance}
%--------------------------------------------------------------

The conductance in the Kondo regime is affected by the barrier asymmetry in the same way as for a resonant tunneling peak in the Coulomb blockade regime, because the conduction in the Kondo regime is a resonant tunneling process locked at the Fermi level. 
The only difference is the factor 2 in the pre-factor coming from spin degeneracy, since Coulomb blockade is lifted in the Kondo regime. 
The zero-bias and zero-temperature conductance is therefore given by \cite{glazman_lowtemperature_2005}
\begin{equation}
    G = \frac{2e^2}{h} 
    \,\frac{4\Gamma_{\rm L}\Gamma_{\rm R}}{(\Gamma_{\rm L}+\Gamma_{\rm R})^2} \, .
    \label{eq:conductance_SM} 
\end{equation}

\begin{figure}[!h]
    \centering
    \includegraphics[width=14cm]{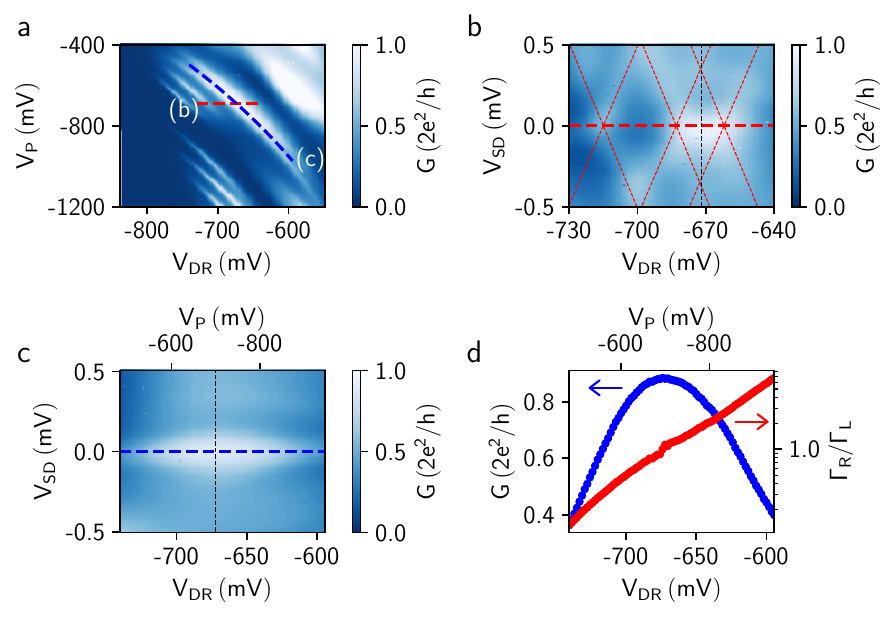}
    \caption{
    \textbf{Evolution of the Kondo conductance with barrier asymmetry.}
    (a) Zero-bias conductance as a function of the gate voltages $V_{\rm P}$ and $V_{\rm DR}$ controlling the left and right barriers, respectively. 
    (b) Energy stability diagram along the red dashed line in (a) with a Kondo ridge in the right Coulomb blockade diamond.
    (c) Evolution of the Kondo resonance in the middle of the Kondo valley along the blue dashed line in (a). The intersection between the maps in (b) and (c) is indicated by the black dashed line.
    (d) Zero-bias conductance along the blue dashed line in (a) and (c). The extracted tunneling rate ratio $\Gamma_{\rm R}/\Gamma_{\rm L}$ is equal to 1 when the conductance reaches its maximum value (symmetric barriers). Note that this maximum equals $0.88\times2e^2/h$ here in absence of cavity, whereas it equals $0.75\times2e^2/h$ in Fig.~2b of the main text due to the presence of the cavity in series with the QD (all parameters are identical in the two figures except the gate voltage $V_{mirror}$ which turns the cavity on and off).
    }
    \label{Barrier_Symmetry}
\end{figure}

In order to analyze the influence of $\Gamma_{\rm L}$ and $\Gamma_{\rm R}$ on the Kondo conductance, we consider the map shown in Fig.~\ref{Barrier_Symmetry}a where the conductance is plotted as a function of the gate voltages $V_{\rm P}$ and $V_{\rm DR}$ controlling essentially the left and right barriers, respectively.
This map has been obtained for the same gate configuration as Fig.~2a of the main text, but without Fabry-Pérot cavity (mirror gate grounded).
The anti-diagonal stripes of large conductance correspond to the Coulomb blockade peaks.

The energy stability diagram obtained by varying $V_{\rm DR}$ along the red dashed line is shown in Fig.~\ref{Barrier_Symmetry}b, with a Kondo resonance in the diamond centered around $V_{\rm DR}=-670$~mV.
The influence of the barrier asymmetry on this Kondo resonance is shown in Fig.~\ref{Barrier_Symmetry}c by varying simultaneously $V_{\rm P}$ and $V_{\rm DR}$ following the blue dashed line in the middle of the Kondo valley.
The zero-bias conductance along this path is plotted in blue in Fig.~\ref{Barrier_Symmetry}d.
At $V_{\rm DR}=-670$~mV and $V_{\rm P}=-700$~mV, it reaches the maximum value $0.88\times 2e^2/h$, closely approaching the unitary limit $2e^2/h$ predicted by Eq.~\eqref{eq:conductance_SM} when $\Gamma_{\rm L}=\Gamma_{\rm R}$. 

Since the increase of $V_{\rm DR}$ increases preferentially the tunneling rate $\Gamma_{\rm R}$ while the simultaneous decrease of $V_{\rm P}$ reduces preferentially $\Gamma_{\rm L}$, there is necessarily a set of gate voltages for which the two tunneling rates are equal and the conductance should be $2e^2/h$.
The observed deviation from a perfect conductance for optimally-symmetrized barriers could have several origins.
It might come from the non-zero temperature of the measurement (50~mK), although much lower than the Kondo temperature (900~mK).
It might also come from decoherence within the QD, although the inelastic scattering time is expected to be much longer than the quantum life time in the QD, in particular in the Kondo regime which requires large tunnel couplings.
One could also wonder if the theoretical model that describes resonant tunneling through a quantum state between two tunnel barriers is fully applicable to the real 2D geometry of a gate-defined QD in a semiconductor heterostructure.
In addition, the QD is connected to 2DEG regions which might not be perfect reservoirs, in particular for the right reservoir where the presence of unused grounded gates might produce a small amount of back-scattering (due to the work function difference between the semiconductor 2DEG and the gate metal).
Finally, there might be some drift in the calibration of the RF measurement setup, resulting in an inaccurate conversion of the RF signal into conductance (the calibration is done at the beginning of the experiment, just after tuning the gate voltages defining the QD).

Because of this deviation from the ideal case, the ratio $\Gamma_{\rm R}/\Gamma_{\rm L}$ is extracted from the zero-bias conductance using Eq.~\eqref{eq:conductance_SM} after normalization to the maximum value $0.88\times 2e^2/h$.
This ratio plotted in red in Fig.~\ref{Barrier_Symmetry}d is therefore equal to 1 at the conductance maximum, and varies by a factor of 10 when the conductance decreases by half.
This analysis is used in the main text to define the two regimes of gate voltages with opposite barrier asymmetry:
\begin{equation*}
\begin{split}
    V_{\rm DR} < -670 \text{ mV and } V_{\rm P} > -700 \text{ mV} \hspace{5mm} \rightarrow \hspace{5mm} \Gamma_{\rm R} < \Gamma_{\rm L} \\
    V_{\rm DR} > -670 \text{ mV and } V_{\rm P} < -700 \text{ mV} \hspace{5mm} \rightarrow \hspace{5mm} \Gamma_{\rm R} > \Gamma_{\rm L}
\end{split}
\end{equation*}

\clearpage
%--------------------------------------------------------------
\subsection{Extraction of the total tunneling rate from the Kondo temperature}
%--------------------------------------------------------------

In the middle of the Kondo valley where $\epsilon_{\rm d}=-U/2$, the Haldane's formula~\eqref{Haldane_full} for the Kondo temperature simplifies to
\begin{equation}
    T_{\rm K} = \frac{\sqrt{\Gamma U}}{2} 
    \,\exp\!\left[-\frac{\pi U}{4 \Gamma}\right] \, .
    \label{TK_Middle_Kondo_Valley3}
\end{equation}
where $\Gamma=\Gamma_{\rm L}+\Gamma_{\rm R}$ is the total tunneling rate (defined as the FWHM of the quantum level) and $U$ the charging energy. 
This expression shows that a stronger coupling $\Gamma$ between the dot and the two leads favors a larger Kondo temperature $T_{\rm K}$. 
It can be used to calculate $\Gamma$ from the measured $T_{\rm K}$ as illustrated below.

Figure~\ref{Sum_Gamma_And_Kondo_Temp}a shows again the Kondo resonance plotted as a function of the linked gate voltages $(V_{\rm DR},V_{\rm P})$ along the blue dashed line in Fig.~\ref{Barrier_Symmetry}a. 
$T_{\rm K}$ is extracted  by fitting the width of the Kondo resonance at each gate voltage.
The scaled curves for each pair $(V_{\rm DR},V_{\rm P})$ are shown in Fig.~\ref{Sum_Gamma_And_Kondo_Temp}b and the extracted Kondo temperatures are plotted in Fig.~\ref{Sum_Gamma_And_Kondo_Temp}c. 

Since the charging energy $U$ is needed to calculate $\Gamma$ from $T_{\rm K}$ using Eq.~\eqref{TK_Middle_Kondo_Valley3}, it should be determined from the height of the Coulomb blockade diamond corresponding to the Kondo valley. 
From Fig.~\ref{Barrier_Symmetry}b, we deduce $U=370~\mu$eV.
The resulting values of the total tunneling rate $\Gamma$ are shown in Fig.~\ref{Sum_Gamma_And_Kondo_Temp}c.

Interestingly, $\Gamma$ is not constant when varying simultaneously the two gate voltages in opposite directions. 
Naively, $\Gamma$ could remain constant because a slight closing of one barrier would be compensated by a slight opening of the other.
However, the gate dependence of the tunneling rate has no reason to be linear, it should instead have an exponential dependence characteristic of a tunnel barrier.
In this case, $\Gamma$ should increase away from the symmetry point, since an increase of one rate would be always larger than the decrease of the other rate. 
The observed decrease of $\Gamma$ away from the symmetry point is therefore not compatible with a fully exponential dependence, and rather indicates a saturating gate dependence, which will be revealed by calculating the two tunneling rates separately, as shown in the next section.

Independently of the above discussion, one should wonder if the formula \eqref{TK_Middle_Kondo_Valley3} (used to extract $T_{\rm K}$ and $\Gamma$) is valid in our case. 
This formula was indeed demonstrated by Haldane \cite{haldane_atomic_1978} for a perturbative coupling $\Gamma \ll U$, which is \textit{a posteriori} not the case here, since we find $\Gamma \sim U$. 
In addition, the formula was demonstrated for a magnetic impurity in a metal, not for a gate-controlled quantum dot with two barriers, such that a possible influence of the barrier asymmetry is not included. 
These restrictions imply that we should not give too much importance to the small variations of $\Gamma$ observed in Fig.~\ref{Sum_Gamma_And_Kondo_Temp}c.

\begin{figure}[!h]
    \centering
    \includegraphics[width=17cm]{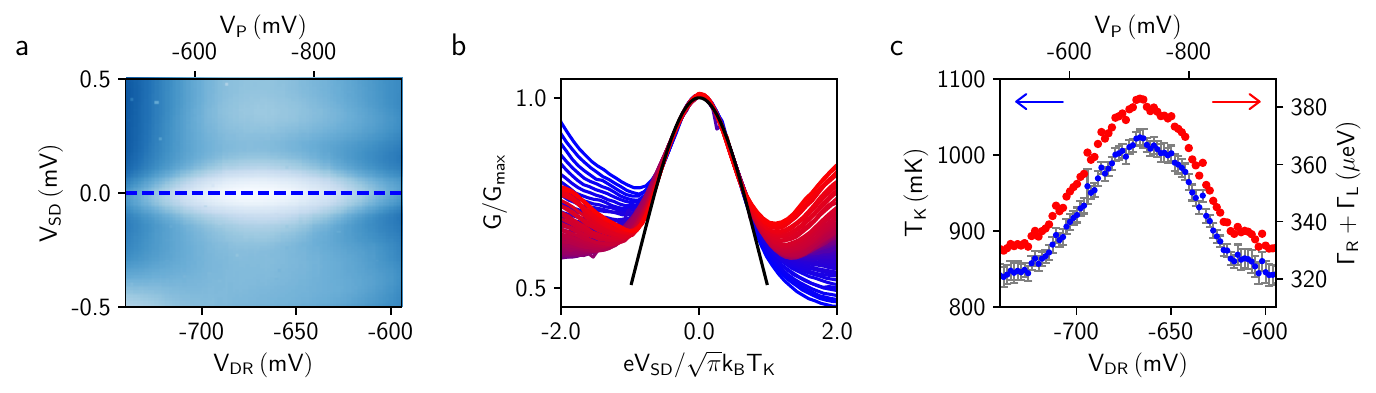}
    \caption{
    \textbf{Evolution of the Kondo temperature with barrier asymmetry.} 
    (a) Evolution of the Kondo resonance along the blue dashed line in Fig.~\ref{Barrier_Symmetry}a. 
    (b) Normalized conductance $G/G_{\rm max}$ as a function of the rescaled bias $eV_{\rm SD}/\sqrt{\pi}k_{\rm B}T_{\rm K}$ along the blue dashed line. 
    (c) Extracted Kondo temperature $T_{\rm K}$ along the blue dashed line and corresponding values of the total tunneling rate $\Gamma$.
    }
    \label{Sum_Gamma_And_Kondo_Temp}
\end{figure}

\clearpage
%--------------------------------------------------------------
\subsection{Extraction of the left and right tunneling rates}
%--------------------------------------------------------------

\begin{figure}[!h]
    \centering
    \includegraphics[width=17cm]{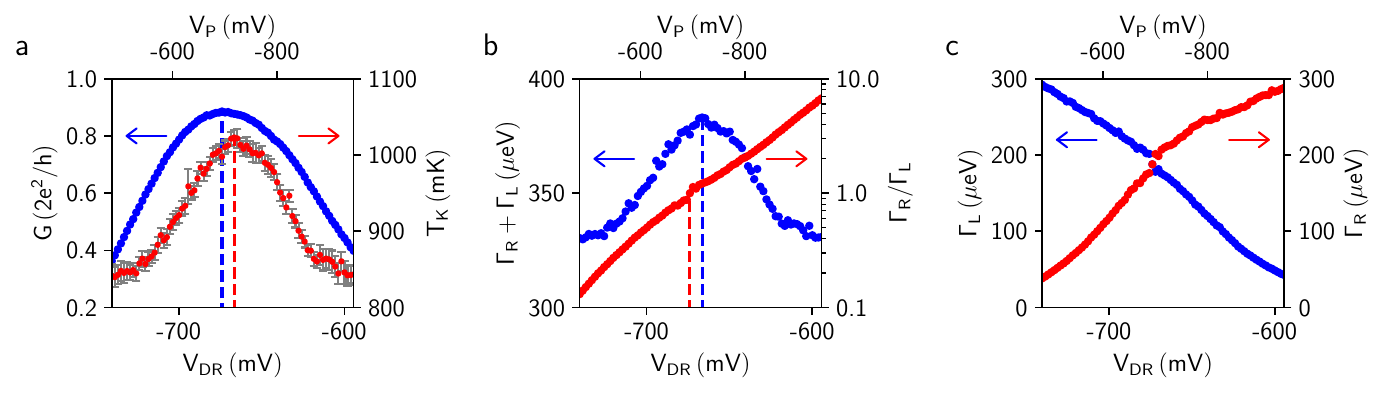}
    \caption{
    \textbf{Extraction of the two tunneling rates.}
    (a) Comparison between Kondo conductance and Kondo temperature along the blue dashed line in Fig.~\ref{Barrier_Symmetry}a. 
    (b) Total tunneling rate $\Gamma=\Gamma_{\rm L}+\Gamma_{\rm R}$ and tunneling rate ratio $\Gamma_{\rm R}/\Gamma_{\rm L}$ calculated from the quantities in (a). 
    (c) Independent evolution of the two tunneling rates with the gate voltages controlling the two barriers.
    }
    \label{Synthese_Gamma_Tk_Ratio_}
\end{figure}

Figure~\ref{Synthese_Gamma_Tk_Ratio_}a compares the evolution of the zero-bias conductance and Kondo temperature (see previous sections) while varying simultaneously $V_{\rm P}$ and $V_{\rm DR}$ to stay in the middle of the Kondo valley.
Interestingly, the maximum of conductance and the maximum of Kondo temperature are obtained at the same setting of gate voltages (the two vertical dashed lines coincide). 
The sum $\Gamma_{\rm L}+\Gamma_{\rm R}$ (extracted from the Kondo temperature) is therefore maximum when the ratio $\Gamma_{\rm R}/\Gamma_{\rm L}$ (extracted from the conductance) is equal to 1, as shown in Fig.~\ref{Synthese_Gamma_Tk_Ratio_}b.

From the ratio and the sum of the two tunneling rates, one can calculate separately $\Gamma_{\rm R}$ and $\Gamma_{\rm L}$, which are plotted in Fig.~\ref{Synthese_Gamma_Tk_Ratio_}c.
Interestingly, the gate dependence of each tunneling rate shows an inflection point, marking the transition from an exponential increase to a saturation (which gives a maximum in the sum).
This saturation might be related to the sigmoid shape of the transmission through the saddle potential of a QPC, with an inflection point at transmission 1/2 when the Fermi level is aligned with the saddle point.

Finally, since each tunneling rate is mainly controlled by only one of the two gate voltages, Fig.~\ref{Synthese_Gamma_Tk_Ratio_}c provides separately the transmission curves $\Gamma_{\rm L}(V_{\rm P})$ for the left QPC and $\Gamma_{\rm R}(V_{\rm DR})$ for the right QPC.
The two transmission curves are very similar, indicating similar potential barriers for the two QPCs, but the lever-arm is three times smaller for $V_{\rm P}$ than for $V_{\rm DR}$ since the plunger gate electrode of the QD is quite far from the left QPC (normally controlled by $V_{\rm UL}$).

\clearpage
%--------------------------------------------------------------
\noindent\hrulefill\vspace{-5mm}
\section{Anomalous temperature dependence of Kondo dots coupled to long cavities}
\vspace{-5mm}\noindent\hrulefill
%--------------------------------------------------------------

\vspace{5mm}
This section discusses the issue of measuring the Kondo temperature from the temperature dependence of the conductance, when the QD is coupled to the Fabry-Pérot cavity which creates interference in the right reservoir.
The objective of this measurement is to observe oscillations of the Kondo temperature resulting from the oscillations of the DOS at the Fermi level in this reservoir.
However, the conductance of the system is found to increase with temperature in the minima of the conductance oscillations, a behavior which is not compatible with the measurement of the Kondo temperature via the decrease of the conductance with temperature.
This behavior results from the thermal broadening of the interference, which is stronger for long cavities with small level spacing.
This effect is evidenced experimentally in the section below, and analyzed theoretically in the subsequent sections.

%--------------------------------------------------------------
\subsection{Temperature dependence of the conductance oscillations (experiment)}
%--------------------------------------------------------------

\begin{figure}[!h]
    \centering
    \includegraphics[width=14cm]{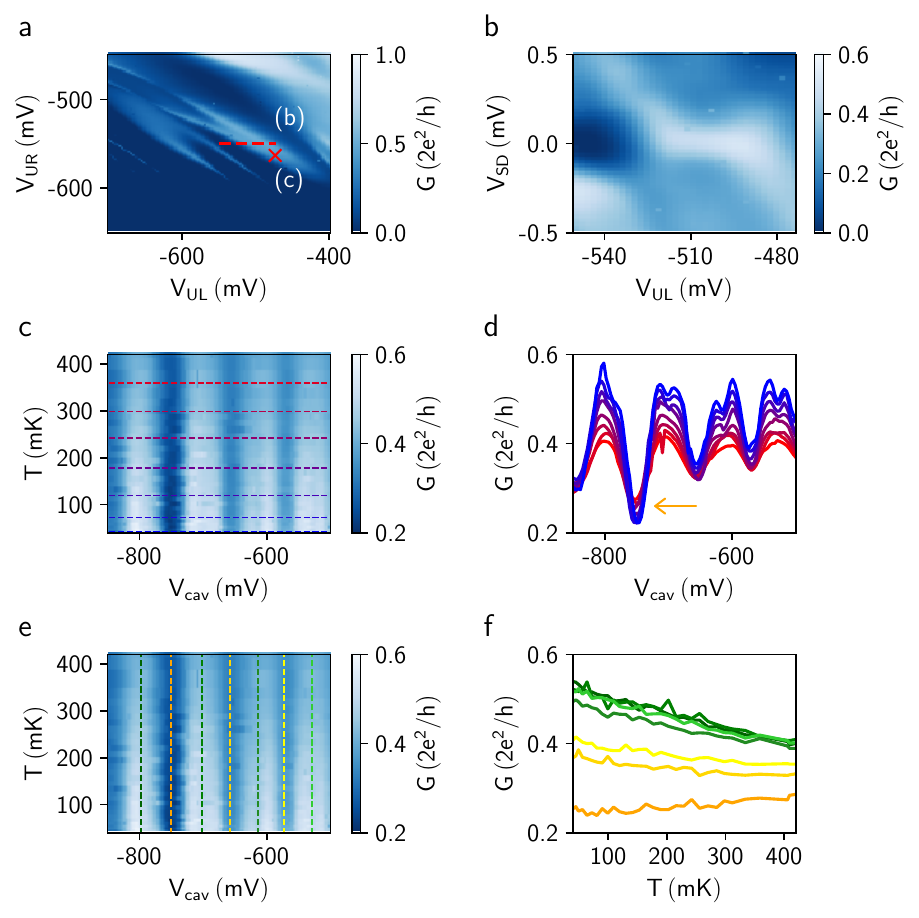}
    \caption{
    \textbf{Evolution of the interference fringes with temperature.}
    (a) Zero-bias conductance as a function of two gate voltages, revealing Coulomb blockade peaks. 
    (b) Energy stability diagram along the red dashed line in (a) without Fabry-Pérot cavity. A Kondo resonance is present in the Coulomb blockade diamond around $-500$~mV. 
    (c) Conductance oscillations as a function of temperature and cavity gate voltage, for fixed QD gate voltages $V_{\rm UL}=-473$~mV and $V_{\rm UR}=-563$~mV in the middle of the Kondo valley (red cross). 
    (d) Conductance oscillations for different temperatures corresponding to the dashed lines in (c).
    }
    \label{TKNeg}
\end{figure}

Figure~\ref{TKNeg}a presents a charge stability diagram as a function of the gate voltages $V_{\rm UR}$ and $V_{\rm UL}$ which tunes both the barrier heights and the dot occupancy. 
The energy stability diagram obtained by sweeping the gate voltage along the red dashed line is shown in Fig.~\ref{TKNeg}b.
A Kondo resonance at zero-bias is visible in one of the Coulomb blockade diamonds. 
The Kondo temperature in the middle of this Kondo valley is about 800~mK. 

Fixing the gate voltages in the middle of this Kondo valley at the position of the red cross in Fig.~\ref{TKNeg}a, the zero-bias conductance is measured as a function of the cavity gate voltage $V_{\rm cav}$ and the temperature $T$ in Fig.~\ref{TKNeg}c. 
The conductance oscillations are plotted in Fig.~\ref{TKNeg}d for different temperatures, in order to compare the temperature dependence of the conductance at maxima and minima of the oscillations.
The conductance decreases strongly with temperature at the maxima and, most of the time, decreases weakly with temperature at the minima. 
This difference corresponds to a higher Kondo temperature at the minima.
However, at the minimum located at $V_{\rm cav}=-750$~mV (arrow), the conductance increases with temperature, preventing the extraction of the Kondo temperature from a fit with a decreasing function. 
This issue led us to extract the Kondo temperature from the bias dependence of the conductance at the base temperature of the fridge, instead of from the usual temperature dependence.

\clearpage
%--------------------------------------------------------------
\subsection{Modeling the conductance of a Kondo dot}
%--------------------------------------------------------------

To understand the anomalous temperature dependence reported in the previous section, we calculate the finite-temperature conductance of a Kondo dot coupled to a Fabry-Pérot cavity using a simple model for the energy-dependent transmission probability through a QD in the Kondo regime.
The conductance at finite temperature is calculated with the general relation
\begin{equation}
    G(T) = \frac{2e^2}{h} \int \left(-\frac{df}{dE}\right) \mathcal{T}(E) \, dE
    \label{Conductance_Temp_Kondo}
\end{equation}
where $f(T,E)$ is the Fermi-Dirac distribution and $\mathcal{T}(E)$ is the transmission at energy $E$ through the coherent system made of a Kondo dot in series with a cavity. 

\begin{figure}[!h]
    \centering
    \includegraphics[width=8cm]{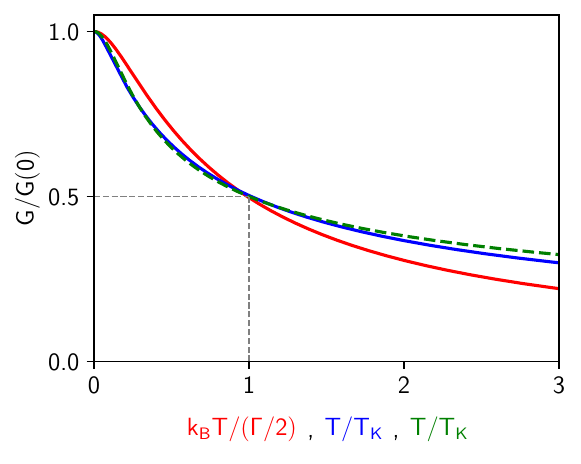}
    \caption{
    \textbf{Temperature dependence of the conductance peak for a quantum level and for a Kondo resonance.}
    Red line: conductance of a quantum level with a Lorentzian transmission given by Eq.~\eqref{Breit_Wigner_Formula}, as a function of the normalized temperature $k_{\rm B}T/(\Gamma/2)$. 
    Blue line: conductance of a Kondo resonance with a transmission given by Eq.~\eqref{Kondo_Transmission}, as a function of the normalized temperature $T/T_{\rm K}$.
    Green dashed line: Kondo conductance given by Eq.~\eqref{Formula_T_Empirical} fitting the exact NRG calculations.
    }
    \label{Breit_WignerVSKondo}
\end{figure}

In absence of cavity and in absence of Kondo effect, the transmission through a single quantum state at energy $E_{\rm res}$ is given by the Lorentzian function \cite{breit_capture_1936}
\begin{equation}
    \mathcal{T}(E) = \frac{4\Gamma_{\rm L}\Gamma_{\rm R}}{\left(\Gamma_{\rm L}+\Gamma_{\rm R}\right)^2} \, \left[\frac{\left(\Gamma/2\right)^2}{\left(\Gamma/2\right)^2+\left(E-E_{\rm res}\right)^2}\right] 
    \label{Breit_Wigner_Formula}
\end{equation}
where $\Gamma=\Gamma_{\rm L}+\Gamma_{\rm R}$ is the FWHM of the resonance, which gives the lifetime $\hbar/\Gamma$ of the localized state. 
The temperature dependence of the conductance peak, when the quantum state is at the Fermi level with $E_{\rm res}=0$, is shown by the red line in Fig.~\ref{Breit_WignerVSKondo}.
This dependence is that of a Coulomb blockade peak in the quantum regime.

In the Kondo regime, the exact temperature dependence of the zero-bias conductance is obtained from NRG calculations and is usually approximated by the phenomenological expression \cite{goldhaber-gordon_kondo_1998}
\begin{equation}
    G(T) = G_{\rm max} \left[ \frac{T_{\rm K}^{'2}}{T_{\rm K}^{'2}+T^2} \right]^{s_2}
    \label{Formula_T_Empirical}
\end{equation}
with $G_{\rm max}=(2e^2/h)(4\Gamma_{\rm L}\Gamma_{\rm R}/\Gamma^2)$, $T_{\rm K}^{'}=T_{\rm K}/\sqrt{2^{1/s_2}-1}$ and $s_2=0.22$ for spin 1/2 electrons.
This temperature dependence is plotted in green in Fig.~\ref{Breit_WignerVSKondo}. 
It differs significantly from the dependence obtained with the Lorentzian transmission \eqref{Breit_Wigner_Formula}.
The transmission $\mathcal{T}(E)$ for a Kondo resonance is indeed not a simple Lorentzian function and it also depends on temperature. 
Unfortunately, there is no analytical expression for the spectral density at finite temperature, that would have provided an expression for $\mathcal{T}(E)$ in the Kondo regime.
As an alternative, we assume an approximate form for $\mathcal{T}(E)$ based on the bias dependence $G(V_{\rm SD})$ of the conductance in the Kondo regime, using Eq.~(3) of Ref.~\cite{kretinin_universal_2012} that fits the exact NRG calculations at finite bias.
In addition, we apply the correspondence $eV_{\rm SD}=2E$ explained in Ref.~\cite{hershfield_probing_1991}.
We therefore write the transmission of the Kondo resonance locked at the Fermi level as
\begin{equation}
    \mathcal{T}(E) = \frac{4\Gamma_{\rm L}\Gamma_{\rm R}}{\left(\Gamma_{\rm L}+\Gamma_{\rm R}\right)^2} \, \left[\frac{(k_{\rm B}T_{\rm K}^{''})^2}{(k_{\rm B}T_{\rm K}^{''})^2+(2E)^2}\right]^{s_1}
    \label{Kondo_Transmission}
\end{equation}
with $T_{\rm K}^{''}=T_{\rm K}\sqrt{\pi/(2^{1/s_1}-1)}$ and $s_1=0.32$. 
Note that this approach is only valid at small bias, when the system is close to equilibrium \cite{gerland_transmission_2000}. 
The temperature dependence of the Kondo resonance is shown by the blue line in Fig.~\ref{Breit_WignerVSKondo}.
This dependence is in good agreement with the green dashed line matching the NRG results.

\clearpage
%--------------------------------------------------------------
\subsection{Modeling the coupling rate of a Kondo dot coupled to a cavity}
%--------------------------------------------------------------

To model the influence of the cavity on the Kondo conductance at finite temperature, we need the energy dependence of the tunneling rate $\Gamma_{\rm R}(E)$.
This quantity can be calculated using the 1D tight-binding model represented in Fig.~\ref{TB_model}, as shown in the Supplementary Information file of Ref.~\cite{borzenets_observation_2020}, based on the theoretical work reported in Ref.~\cite{park_measure_2013}.
The dot is located at site $n=0$ with an onsite energy $\epsilon_{\rm d}$. 
It is coupled to the left (L) and right (R) leads by the hopping energies $V_{\rm L}$ and $V_{\rm R}$, which are much smaller than the hopping energy $t$ between the sites of the leads.
The dispersion relation for an electron of energy $E$ and wave vector $k$ is given by $E=-2t\cos(ka)$ where $a$ is the lattice parameter.
To account for the presence of the FP cavity of size $L=Na$ on the right side of the dot, the hopping parameter between site $N$ and $N+1$ is changed to $t_0$ smaller than $t$. 
This inhomogeneity in the chain scatters the electron wave and simulates the reflection induced by the cavity mirror.

\begin{figure}[!h]
    \centering
    \includegraphics[width=13cm]{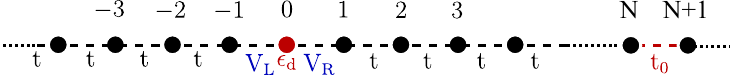}
    \caption{
    \textbf{1D tight-binding model of a QD coupled to a FP cavity.}
    The dot is located at site $n=0$ with an onsite energy $\epsilon_{\rm d}$. The parameters $V_{\rm L}$ and $V_{\rm R}$ are the hopping energies between the dot and the lead sites $n=-1$ and $n=1$, respectively. The rest of the chain is composed of sites with an onsite energy fixed to zero and coupled to each other with a hopping energy $t$, except the sites $N$ and $N+1$ which are coupled with a hopping energy $t_0$ to model the cavity mirror.
    }
    \label{TB_model}
\end{figure}

This model has been used in Ref.~\cite{borzenets_observation_2020} to calculate the tunneling rate $\Gamma_{\rm R}(E)$ given in Eq.~(S3) of their Supplementary Information file, and expressed as 
\begin{equation} 
    \Gamma_{\rm R}(E) = \frac{\Gamma_{\rm R\infty}}
    {(1-\alpha)\cos^2\!\left(\frac{\pi E}{\Delta}+k_{\rm F}L\right)
    +\frac{1}{1-\alpha}\,\sin^2\!\left(\frac{\pi E}{\Delta}+k_{\rm F}L\right)}
    \label{Gamma_Evolution}
\end{equation}
where $\Gamma_{\rm R\infty}=2V_{\rm R}^2/t$ is the tunneling rate in absence of cavity, $\alpha=1-(t_0/t)^2$ characterizes the strength of the reflection by the cavity mirror, $\Delta=2\pi ta/L$ is the level spacing in the cavity, $E$ is the electron energy relative to the Fermi level set at $E_{\rm F}=0$, and $k_{\rm F}=\pi/2a$ is the Fermi wave vector.
Note that $\Gamma$ is defined in our work as the full width at half maximum (FWHM) of the quantum level, like in the early experimental works \cite{PhysRevLett.81.5225,vandewiel_unitary_2000}, while it is defined as the half width at half maximum (HWHM) in \cite{park_measure_2013,borzenets_observation_2020} and in most theoretical works.
More details about this model are also given in \cite{fossion_phd_2024}.

In the discussion of the main text where the conductance is measured at very low temperature in the regime $T\ll T_{\rm K}$, the relevant tunneling rate is the value at the Fermi level. 
$\Gamma_{\rm R}$ is therefore obtained from Eq.~\eqref{Gamma_Evolution} by setting $E=0$ and is further simplified by assuming a small parameter $\alpha\ll1$ corresponding to a small contrast of the interference fringes, which leads to
\begin{equation}
    \Gamma_{\rm R} = \Gamma_{\rm R\infty} \, \big(1+\alpha\cos(2k_{\rm F}L)\big)
    \label{GammaR_simplified}
\end{equation}
This expression is also shown as Eq.~(3) of the main text and is used to plot the graphs of Fig.~4 for $U=0.1t$, $\Gamma=0.06t$, $t_0=0.95t$ ($\alpha=0.0975$).

\clearpage
%--------------------------------------------------------------
\subsection{Temperature dependence of the conductance oscillations (model)}
%--------------------------------------------------------------

The conductance of the Kondo dot coupled to the cavity at finite temperature is calculated with Eq.~\eqref{Conductance_Temp_Kondo} by inserting the energy- and length-dependent tunneling rate~\eqref{Gamma_Evolution} in the pre-factor of the Kondo transmission~\eqref{Kondo_Transmission}, while the Kondo temperature $T_{\rm K}$ in the spectral function only depends on the cavity length. 

The calculation is done for a given set of parameters $\Gamma_{\rm L}$ and $\Gamma_{\rm R\infty}$.
These parameters control the Kondo temperature $T_{\rm K\infty}$ and the corresponding Kondo length $\xi_{\rm K\infty}=\hbar v_{\rm F}/k_{\rm B}T_{\rm K\infty}$.
The tunneling rates are chosen such that $\xi_{\rm K\infty}=35\lambda_{\rm F}$ in order to reproduce the conditions of the experiment where $T_{\rm K}\sim900$~mK corresponds to $\xi_{\rm K}\sim1.75~\mu\mathrm{m}=35\lambda_{\rm F}$.
Since the FP cavity has a length $L=3.2~\mu\mathrm{m}=64\lambda_{\rm F}$, the experiment is in the regime $L>\xi_{\rm K}$.

\begin{figure}[!h]
    \centering
    \includegraphics[width=14cm]{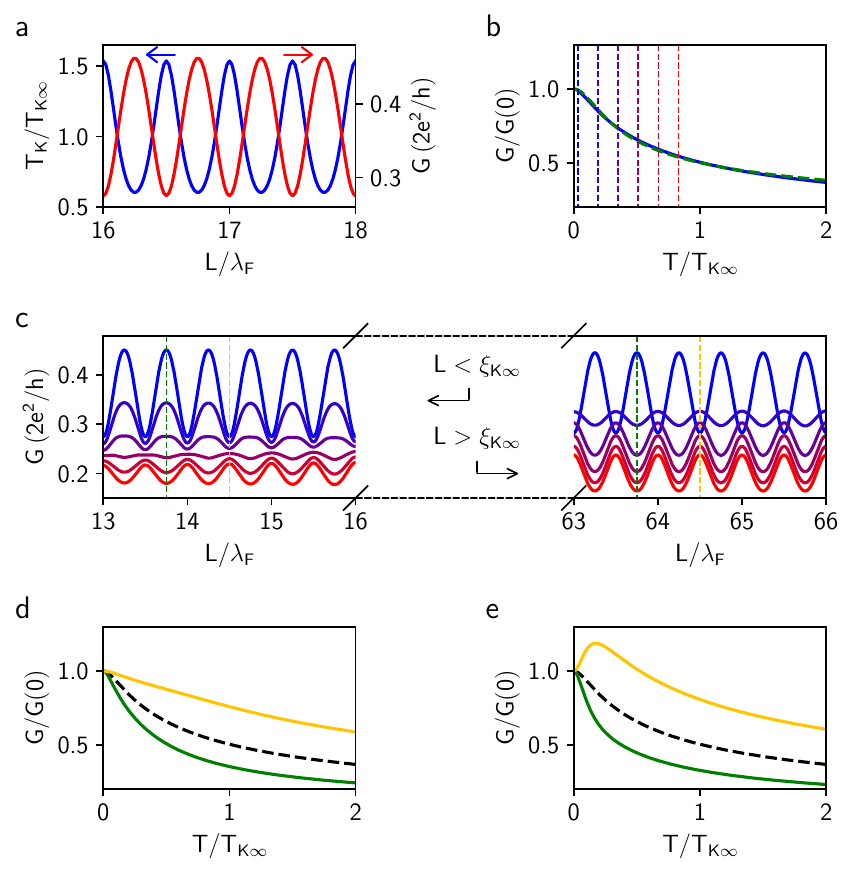}
    \caption{
    \textbf{Temperature dependence of the conductance in a Kondo dot coupled to a Fabry-Pérot interferometer.}
    (a) Oscillations of the normalized Kondo temperature in blue and of the zero-temperature conductance in red for $\Gamma_{\rm R\infty}/\Gamma_{\rm L}=9$.
    (b) Normalized conductance as a function of scaled temperature $T/T_{\rm K\infty}$ in absence of cavity ($t_0=t$) calculated with the spectral function~\eqref{Kondo_Transmission} (blue solid line) and with the phenomenological expression~\eqref{Formula_T_Empirical} (green dashed line).
    (c) Oscillations as a function of cavity length in the regimes $L<\xi_{\rm K\infty}$ (left part) and $L>\xi_{\rm K\infty}$ (right part) for the temperatures indicated in panel (b) by vertical dashed lines.
    (d,e) Evolution of the normalized conductance as a function of the normalized temperature in the two regimes of cavity length. The curves in green and yellow correspond to maxima and minima of the zero-temperature conductance oscillations, respectively. The black dashed line is the normalized conductance in absence of cavity. 
    The calculations have been done for $\epsilon_{\rm d}=-U/2$, $U=0.1t$, $V_{\rm L}=0.06t$, $V_{\rm R}=0.18t$, and $t_0=0.85t$.
    }
    \label{Calc_TK}
\end{figure}

In addition, we adjust the two tunneling rates to have a ratio $\Gamma_{\rm R\infty}/\Gamma_{\rm L}=9$, corresponding to a regime of barrier asymmetry with a dot more coupled to the cavity.
The zero-temperature conductance and Kondo temperature calculated with Eqs.~(1) and (2) of the main text are both plotted in Fig.~\ref{Calc_TK}a, exhibiting out-of-phase oscillations as in Fig.~4c of the main text.

The influence of temperature on conductance oscillations is plotted in Fig.~\ref{Calc_TK}c, for a series of temperatures in the regime $T<T_{\rm K\infty}$ where the Kondo conductance decreases rapidly with temperature (see Fig.~\ref{Calc_TK}b in absence of cavity).
The amplitude of the oscillations also decreases with temperature, not only on the energy scale $T_{\rm K\infty}$ of the Kondo effect, but also on the energy scale of the cavity level spacing $\Delta$. 

In the regime $L<\xi_{\rm K\infty}$ shown in the left part of Fig.~\ref{Calc_TK}c, the oscillation amplitude decreases slowly with temperature, and the temperature dependence of the conductance is monotonous both in maxima and in minima of the oscillations, as illustrated in Fig.~\ref{Calc_TK}d. 
In this regime, the Kondo temperature extracted using the empirical relation~\eqref{Formula_T_Empirical} would be higher for the green curve corresponding to a minimum of conductance, as a result of the chosen barrier asymmetry with a dot more coupled to the cavity (regime with out-of-phase oscillations).
Interestingly, an inversion of the conductance oscillations appears at high temperature due to a very different temperature dependence on maxima and minima of the zero-temperature oscillations.

On the other hand, in the regime $L>\xi_{\rm K\infty}$ shown in the right part of Fig.~\ref{Calc_TK}c, the effect of temperature on the conductance oscillations is more pronounced at low temperature, because of a stronger thermal broadening of the DOS oscillations in a long cavity with $\Delta<k_{\rm B}T_{\rm K\infty}$.
This strong attenuation of the oscillation amplitude gives rise to a non-monotonous temperature dependence of the conductance at the oscillation minima, as shown by the green curve in Fig.~\ref{Calc_TK}e.
Such a dependence cannot be fitted with the empirical relation~\eqref{Formula_T_Empirical} of the Kondo effect.

In conclusion, the temperature dependence of the Kondo conductance cannot be used reliably to extract the Kondo temperature oscillations as a function of the cavity length $L$, in particular in the regime $L>\xi_{\rm K}$.
In our device, the cavity length $L=3.2~\mu$m is larger than the Kondo length $\xi_{\rm K}\sim1.75~\mu$m, such that our experiment falls within this regime.
An example of anomalous temperature dependence is shown in Fig.~\ref{TKNeg}d. 
In our study, we therefore employed an alternative method to extract the Kondo temperature, by measuring the width of the Kondo resonance as a function of bias at very low temperature.

\clearpage
%--------------------------------------------------------------
\noindent\hrulefill\vspace{-5mm}
\section{Conductance and Kondo temperature oscillations for several barrier asymmetries}
\vspace{-5mm}\noindent\hrulefill
%--------------------------------------------------------------

\begin{figure}[!h]
    \centering
    \includegraphics[width=9.3cm]{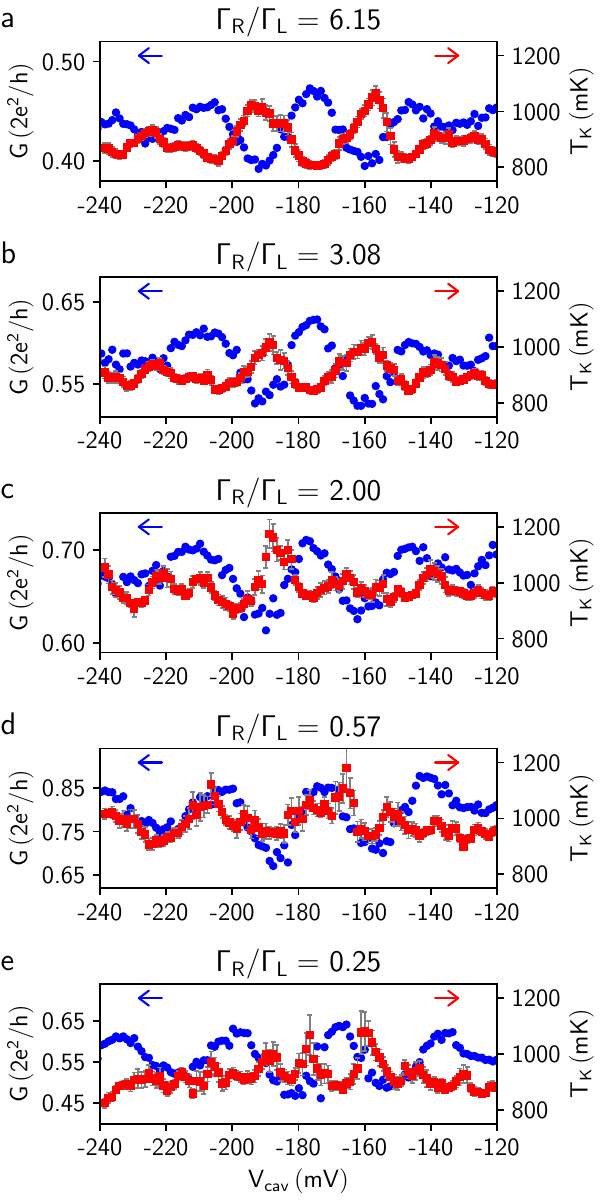}
    \caption{
    \textbf{Conductance and Kondo temperature oscillations for several barrier asymmetries $\Gamma_{\rm R}/\Gamma_{\rm L}$.}
    (a,d) and (b,c,e) correspond the gate voltage configurations indicated by yellow and grey dots in Fig.~2a of the main text, respectively. The graphs in (a) and (d) correspond to Fig.~2c and 2d, respectively.
    }
    \label{Calc_TKk}
\end{figure}

\clearpage
%--------------------------------------------------------------
\noindent\hrulefill\vspace{-5mm}
\section{Theoretical analysis of the oscillation amplitude versus barrier asymmetry}
\vspace{-5mm}\noindent\hrulefill
%--------------------------------------------------------------

\vspace{5mm}
The oscillations of the conductance and Kondo temperature can be computed analytically in the regime of small interference contrast, where the tunneling rate to the cavity is given by Eq.~(3) of the main text with $\alpha\ll1$. 
Writing $x=\Gamma_{\rm R\infty}/\Gamma_{\rm L}$, the conductance and Kondo temperature given by Eqs.~(1) and (2) of the main text can be written as
\begin{eqnarray}
    G &\approx& G_\infty \left[1+\left(\frac{1-x}{1+x}\right)\alpha\cos(2k_{\rm F}L)\right] \\
    T_{\rm K} &\approx& T_{\rm K\infty} \left[1+\left(\frac{x}{1+x}\right)\beta\alpha\cos(2k_{\rm F}L)\right]
\end{eqnarray}
with $\beta=\frac{1}{2}+\frac{\pi U}{4\Gamma_\infty}$. 
The conductance oscillations have a relative amplitude changing sign at $x=1$ and saturating to $\alpha$ at large asymmetry, while the Kondo temperature oscillations have a relative amplitude increasing progressively with $x$ and saturating to $\beta\alpha$ at large asymmetry.
These evolutions with the tunneling rate ratio are visible in Fig.~4 of the main text.

\clearpage
%--------------------------------------------------------------
\noindent\hrulefill\vspace{-5mm}
\section{Gate voltage configuration for measurements of the main figures}
\vspace{-5mm}\noindent\hrulefill
%--------------------------------------------------------------

\begin{table}[!h]
\begin{tabular}{|c|c|c|c|c|c|c|c|}
\hline
Figure & $V_{\rm DL}$ & $V_{\rm UR}$ & $V_{\rm P}$ & $V_{\rm UL}$ & $V_{\rm DR}$ & $V_{\rm cav}$ & $V_{\rm mirror}$ \\
\hline
\hline
1c & -1000 & -550 & -460 & -650 & $\longleftrightarrow$ & 0 & -800 \\
\hline
1d & -1000 & -550 & -460 & -670 & -683 & $\longleftrightarrow$ & -800 \\
\hline
\hline
2a & -1000 & -530 & $\longleftrightarrow$ & -520 & $\longleftrightarrow$ & 0 & -800 \\
\hline
2c & -1000 & -530 & -950 & -520 & -596 & $\longleftrightarrow$ & -800 \\
\hline
2d & -1000 & -530 & -600 & -520 & -700 & $\longleftrightarrow$ & -800 \\
\hline
\hline
3a & -1000 & -550 & -460 & $\longleftrightarrow$ & $\longleftrightarrow$ & 0 & -800 \\
\hline
3b & -1000 & -550 & -460 & $\longleftrightarrow$ & $\longleftrightarrow$ & $\longleftrightarrow$ & -800 \\
\hline
\end{tabular}
\caption{Gate voltages in mV for measurements shown in the main figures. 
The symbol $\longleftrightarrow$ means that the voltage is swept during the measurement.}
\end{table}

\begin{figure}[!h]
    \centering
    \includegraphics[width=10cm]{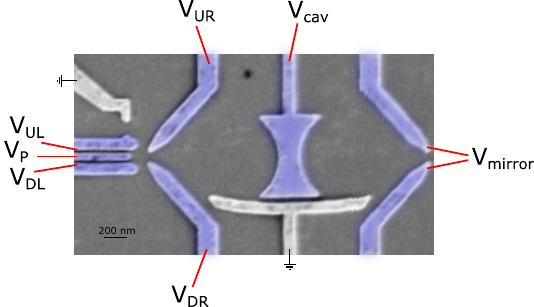}
    \caption{
    Gate voltage names and scanning electron microscope image of a device identical to the one studied in this work.
    }
    \label{SEM_image}
\end{figure}

\clearpage
%--------------------------------------------------------------

%--------------------------------------------------------------

%--------------------------------------------------------------